\documentclass[letterpaper,twocolumn,amsmath,amssymb]{revtex4}
\usepackage[charter]{mathdesign}
\usepackage{siunitx}
\usepackage{graphicx}
\usepackage{float}
\usepackage{booktabs}
\usepackage{setspace}
\usepackage[unicode,colorlinks=true,citecolor=blue,urlcolor=blue,linkcolor=blue]{hyperref}
\setcitestyle{numbers,comma,sort&compress,super}
\usepackage{url}
\usepackage[normalem]{ulem}
\usepackage[nameinlink,capitalize]{cleveref}

\usepackage{bbding}

\usepackage{ragged2e}
\newcommand{\ztitle}[1]{\title{\vspace*{3ex} \raggedright\fontfamily{mathfont}\Large\bf {#1}}}
\newcommand{\zauth}[1]{\author{\vspace{6ex}\hspace*{-50em} \RaggedRight\fontfamily{mathfont}\large\bf {#1}}}
\newcommand{\zaffi}[1]{\affiliation{\vspace{1ex} \raggedright\fontfamily{mathfont} {#1}}}
\begin{document}
\ztitle{High temperature superconductivity arising in a metal sheet full of holes}
\zauth{N. Zen}
\zaffi{Device Technology Research Institute, National Institute of Advanced Industrial Science and Technology,\\Tsukuba Central 2-10, Ibaraki 305-8568, Japan\vspace{1ex}\\
  {\it E-mail address:} {\tt\small n.zen@aist.go.jp}\\
  {\it ORCID ID:} \href{https://orcid.org/0000-0003-4897-478X}{\tt\small 0000-0003-4897-478X}
\vspace{1ex}}
\maketitle
\onecolumngrid
\vspace{-1.5ex}
\hrule
\vspace{1.5ex}
\noindent
{\large\bf Abstract}\vspace{1ex}\\
\textit{\textbf{Background:}} By drilling periodic thru-holes in a suspended film, the phonon system can be modified.\vspace{0.5ex}\\
\textit{\textbf{Method:}} Being motivated by the BCS theory, the technique, so-called phonon engineering, was applied to a niobium sheet.\vspace{0.5ex}\\
\textit{\textbf{Results:}} The newly emergent high-$T_{c}$ superconductivity, however, cannot be accounted for by the BCS theory. Rather, its exposed configuration, namely a square-lattice oxygen network, is reminiscent of the copper--oxygen plane in cuprate high-$T_{c}$ superconductors.\vspace{0.5ex}\\
\textit{\textbf{Conclusions:}} It turns out that its magnetic result is consistent with the principle of \emph{giant atom}, which was developed by the forgotten heroes of superconductivity, Fritz London and John Slater, in the 1930s, several decades earlier than the propagation of BCS theory. The superconducting transition feature is discussed on the basis of a comprehensive theory of the \emph{giant atom}---the theory of \emph{hole superconductivity}.\vspace{-0.5ex}\\
\hrule
\vspace{2.5ex}
\noindent
{\it Keywords:}~{\tt\small Metal{\footnotesize~}sheet, Phonon{\footnotesize~}engineering, High{\footnotesize~}$T_{c}$, Meissner{\footnotesize~}effect, Giant{\footnotesize~}atom, Hole{\footnotesize~}superconductivity}
\vspace{13ex}
\begin{center}
\uline{{\large\bf Table of Contents}}\\
\end{center}
\twocolumngrid
\mbox{}
\vspace{-3ex}
\begin{figure}[h!]
\flushright
\includegraphics[width=75mm]{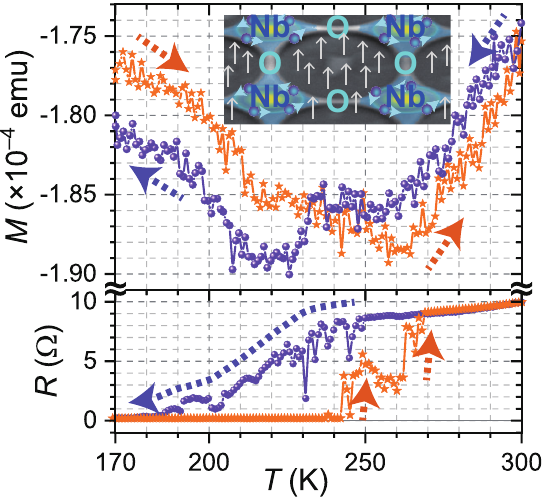}
\end{figure}
\newpage
\mbox{}

\noindent
To claim the occurrence of superconductivity, it is mandatory to show (i) resistance drop to zero, (ii) magnetization drop upon cooling, i.e. the Meissner effect and (iii) the crystal structure. Each result is shown on the bottom panel, upper panel and inset, respectively. All results in this paper were obtained directly from raw data without any background subtraction.
\mbox{}
\newpage
\clearpage
\section{Introduction}
\label{sec:intro}
\begin{figure}[h!]
\centering
\includegraphics[width=75mm]{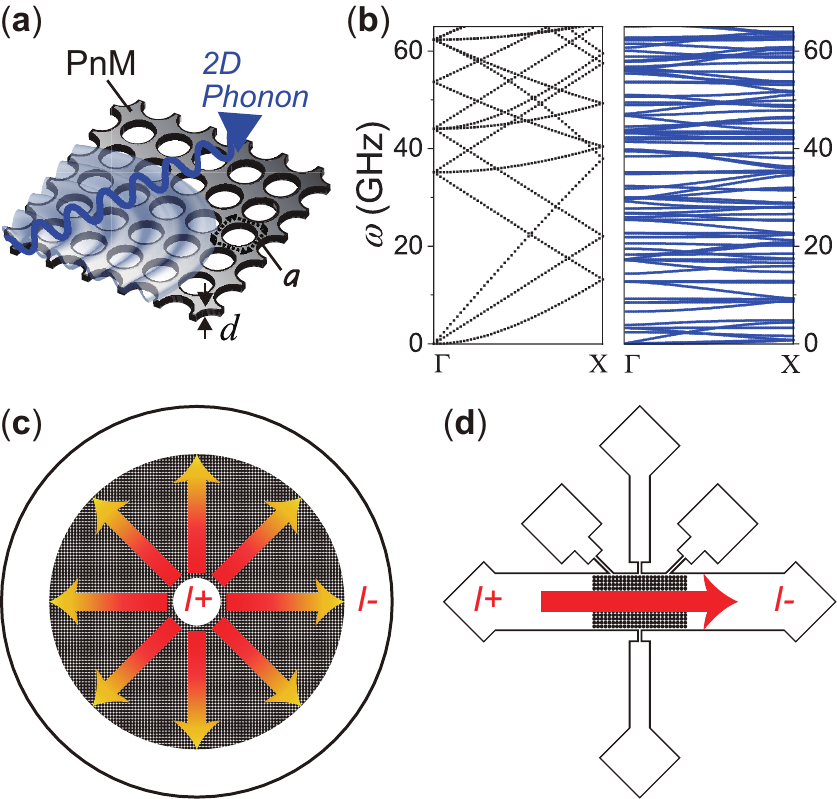}
\caption{\textbf{Phonon engineered metallic system (PnM).} (\textbf{a}) Schematic representation of a 2D PnM. The thickness and lattice constant is indicated by $d$ and $a$, respectively. (\textbf{b}) Phonon dispersions of an Nb sheet with $d$ of 150 nm. (Left) Without perforation. (Right) With periodic perforation designed as the PnM with $a$ of 20 $\si{\micro m}$. (\textbf{c}) and (\textbf{d}) Bird's-eye view of the sample, respectively, used for the previous study~\cite{Zen2019} and for this study.
\label{fig1}}
\end{figure}

According to the BCS theory~\cite{BCS1957}, coherent phonons play a key role in coupling electrons. In other words, a critical temperature $T_{c}$ of superconducting transition is closely associated with the property of coherent phonons, i.e., the phonon dispersion relation. It is generally supposed that the phonon dispersion is material dependent and hence cannot be modified. But it becomes possible, simply by drilling periodic thru-holes in the material (\cref{fig1}a). The periodic perforation induces the elastic analogue of Bragg's interference and therefore affects the coherence of phonon propagation. The technique, known as phonon engineering~\cite{Zen2014}, has attracted physicists and material scientists' attention because of its ability to control thermal properties of dielectric materials~\cite{Mal2015,Soto2019,Nom2020,JAP2021,Pekola2021}. On the other hand, another possibility of the phonon engineering, whether the artificially modified phonon system affects the electron system or not, does not yet get attention.

In the previous study~\cite{Zen2019}, the phonon engineering was applied to a metallic system for the first time. The used material was pure niobium (Nb), the well known conventional superconductor with $T_{c}$ of 9 K. Expecting a change in the $T_{c}$, hopefully to be increased, a pure Nb film with a thickness ($d$) of 150 nm was periodically perforated to form a two-dimensional (2D) square lattice with a lattice constant ($a$) of 20 $\si{\micro m}$. The calculated phonon dispersion of the engineered Nb sheet using these $d$ and $a$ is shown on the right panel of \cref{fig1}b. The left panel shows that of an Nb sheet with the same $d$ without perforation. Obviously overall phonon bands are forcibly flattened by the phonon engineering, and hence the change in $T_{c}$ could be expected. Despite the expectation, however, the $T_{c}$ neither increased nor decreased. Alternatively, the engineered Nb sheet underwent a metal--insulator transition at 43 K during temperature cycles in the temperature range of 2--300 K repeatedly applied to the sample. Independently performed resistive and magnetic measurements revealed that the metal--insulator transition was caused by the 2D Anderson localization of the electron system.

The `2D', despite the film thickness of 150 nm being fully 3D for conducting electrons with a mean free path of a several nm, was the indication of 2D phonon--electron interactions taking place in the sample. A reciprocal lattice in momentum space of the real-space 2D square lattice is also a 2D square lattice. Given that the 2D phonon engineering creates a 2D phonon having a momentum $\vec{q}=(q_{x}, q_{y}, 0)$ and that the 2D phonon interacts with a 3D electron having a momentum $\vec{p}=(p_{x}, p_{y}, p_{z})$, the resultant electron--phonon interaction $\vec{p}\cdot \vec{q}$ is of course 2D. That is, the 3D component of electron momentum is nullified in the 2D phonon space. In other words the 2D phonon engineering may activate the 2D electron system.

On the other hand, the `Anderson localization' was the indication of spatially disordered charge distribution in the sample. In the previous study, the sample had a Corbino disk shape (\cref{fig1}c). That is, an excitation current, which was applied to the sample to measure its electrical resistance, flowed through the sample radially from the center to the periphery reducing its own current density. Accompanied by the emergent 2D phonon--electron interaction during the temperature cycles, the spatially varying current density may cause a disordered charge distribution in the sample.

Ever since the discovery of cuprate superconductors in 1986--87~\cite{Bed1986,Ashburn1987,Muller1987}, especially after the discovery of YBCO~\cite{Ashburn1987} with the $T_{c}$ exceeding the boiling point of liquid nitrogen (77 K), the numerous long-standing research efforts on high temperature superconductivity have been made, and the superconductive community has reached a consensus that the copper--oxygen plane associated with the 2D electron system plays a key role in giving rise to a high $T_{c}$ of cuprate superconductors~\cite{Chu:BCS}. By contrast, the study of 2D phonon-engineered metallic system (PnM) has just begun~\cite{Zen2014,Zen2019}, and regrettably its 2D phonon--electron interaction is not well understood yet. Also the initial expectation, simply motivated by the BCS theory, whether the simple modification of phonon dispersion affects $T_{c}$ or not, seems completely way off the mark. Nevertheless, encouraged by the above consensus on high temperature superconductivity, namely a key role of the 2D electron system, the study of 2D PnM has been continued. In this study, a rectangle shape instead of Corbino is applied to the sample (\cref{fig1}d), wherein an excitation current uniformly flows in a single direction, expecting the Anderson localization to be suppressed and hopefully the $T_{c}$ to be increased.

\section{Material and Methods}
\label{sec:matmet}
\subsection{Material}
\label{sec:mat}
Being different to all other superconducting materials so far reported regardless of whether conventional or unconventional, the sample used in this study was prepared by microfabrication techniques such as sputter deposition, lithography and chemical dry etching. More significantly, the 2D PnM has to be prepared as a self-standing structure. Even if a sample has a periodic structure designed as the PnM, it never shows an anomalous phase transition if it is not self standing. Conversely, even if a sample is self standing, a mere self-standing Nb bridge for instance, which is not designed as the PnM, never shows an anomalous phase transition. Only the 2D PnM sample prepared as a self-standing structure does exhibit an anomalous phase transition, regardless of whether the superconducting transition presented in this study or the insulating transition in the previous study. This fact was insistently investigated previously and is clearly stated in the previous paper~\cite{Zen2019}.

First, a silicon dioxide (SiO\textsubscript{2}) sacrificial layer of 1.0-$\si{\micro m}$ thickness was deposited on a p-type silicon (Si) wafer having the thickness, diameter, orientation, respectively 0.4 mm, 3 inch ($\approx$ 76 mm), (100) by chemical vapor deposition (PD-270STL, Samco) with the stage temperature kept at 80 \si{\degreeCelsius}. The pressure of the mixture of gases of TEOS (tetraethoxysilane) and O\textsubscript{2} was 30 Pa, and the total deposition time was 42 minutes. Post deposition, an Nb film of 150-nm thickness was deposited on the SiO\textsubscript{2} layer by sputtering (M12-0130, Science Plus) at 10 \si{\degreeCelsius}, using Ar gas at 1.0 Pa, for 130 s. Subsequently, an i-line chemical resist (PFi-245, Sumitomo Chemical) was spin-coated to be a thickness of 300 nm on the Nb layer, and the sample patterning was performed to form the square lattice with the lattice constant $a$ of 20 $\si{\micro m}$ using an i-line stepper (NSR-2205i12D, Nikon TEC) with an exposure time of 350 ms. The exposed region of the Nb layer was removed by reactive ion etching (RIE-10NR, Samco) using SF\textsubscript{6} gas at 10.0 Pa for the total etching time of 210 s. After a protective chemical resist (PFi-68A7, Sumitomo Chemical) was spin-coated, the resulting wafer was cut into 5-mm squares using a dicing machine (DAD522, DISCO), and the protective chemical resist was removed. Finally, the SiO\textsubscript{2} sacrificial layer under the already patterned Nb layer was removed by an HF dry etcher (memsstar\textsuperscript{\circledR}SVR\textsuperscript{TM} vHF, Canon). The diced samples were exposed under the mixture of 250-sccm HF gas and water vapor consisting of 100-sccm N\textsubscript{2} and 10-mg H\textsubscript{2}O, with the stage temperature kept at 5 \si{\degreeCelsius}, for the duration of 120 s and 360 s, respectively, for the 8-Torr step and subsequent 9-Torr step. The self-standing PnM-Nb structure was inspected using a laser microscope (LEXT OLS4000, Olympus).

The above microfabrication technique is the same as that used in the previous study~\cite{Zen2019}. That is, the fundamental geometry of the PnM-Nb, namely the thickness $d$ (150 nm), square lattice structure and lattice constant $a$ (20 $\si{\micro m}$), is the same as that of the previous one. Only the shape is different as mentioned in the former section and is rectangle. The GDSII patterning file used for the i-line lithography was deposited at Zenodo~\cite{Zenodo2022} and is freely available.

\begin{figure}[h!]
\centering
\includegraphics[width=70mm]{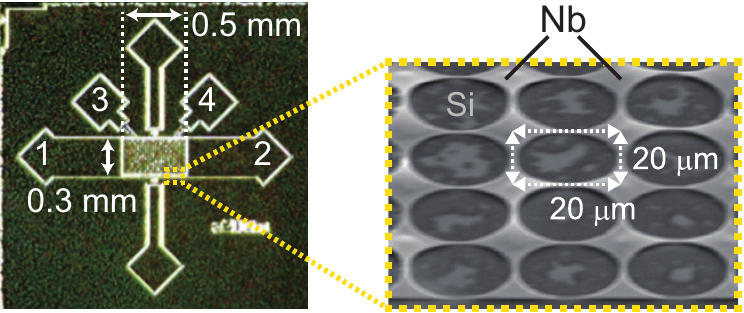}
\caption{Optical micrograph of the sample. Electrical pads are numbered. Right panel, an SEM of the region surrounded by the yellow dotted square.
\label{fig2}}
\end{figure}

\autoref{fig2} shows an optical micrograph of the PnM-Nb sample, and its scanning electron micrograph (SEM) is shown on the right panel. The sample has a rectangle shape with an area of 0.3$\times$0.5 mm\textsuperscript{2}. The Nb sheet with $d$ of 150 nm has the square lattice structure with $a$ of 20 $\si{\micro m}$ and is self standing, approximately 1 $\si{\micro m}$ apart from the Si substrate underneath. The Si substrate is massive, 5$\times$5 mm\textsuperscript{2} in size and 0.4 mm in thickness; its rigidity is necessary to handle the tiny sample.

\begin{figure}[h!]
\centering
\includegraphics[width=0.87\linewidth]{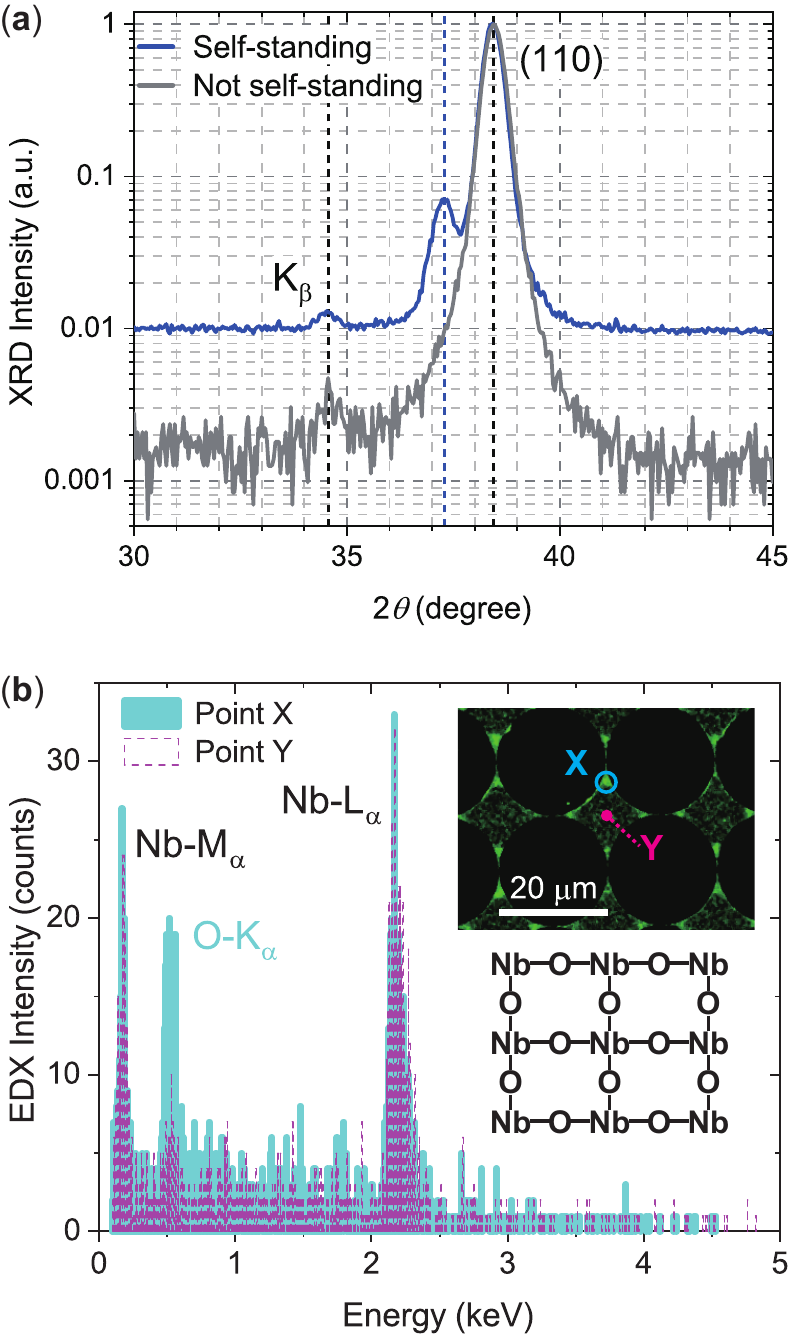}
\caption{\textbf{Structural properties of the sample.} (\textbf{a}) XRD spectra of the self-standing sample and not-self-standing one. (\textbf{b}) EDX spectra of the self-standing sample, where the points ``X'' and ``Y'' are indicated in the SEM-EDX result (inset). Bottom of the inset, a schematic configuration of the SEM-EDX result.
\label{fig3}}
\end{figure}

\autoref{fig3}a shows x-ray diffraction (XRD) spectra of the self-standing sample and a not-self-standing one which was prepared by skipping the final SiO\textsubscript{2} removal procedure. The XRD spectrum of the self-standing sample exhibits an extra peak at 37.3 degree, which is absent for the not-self-standing one. The extra peak corresponds to the atomic lattice spacing of 3.4 \AA, indicating that the self-standing sample contains anomalous regions where the Nb lattice spacing is expanded, a little wider than the usual 3.3 \AA~that can be seen from the main peak at 38.3 degree. The unexpected expansion of lattice spacing might be due to the removal of the SiO\textsubscript{2} sacrificial layer under the Nb layer, which may release an in-plane stress of the Nb layer accumulated during the Nb sputtering process. As discussed later, the expansion seems to have a significant role in giving rise to a high $T_{c}$.

The expansion brings another serendipity. It is well known that light elements such as H, C, N, O in the surrounding atmosphere easily invade a metal. The wider the spacing is, the higher amounts invade. \autoref{fig3}b shows energy dispersive x-ray (EDX) spectra of the self-standing sample, together with its SEM-EDX result in the inset. The accelerating voltage and probe current was 5 kV and 1 nA, respectively, and the integration time was 30 minutes. Obviously, the region pointed as ``X'' includes higher amounts of oxygen than that of ``Y''. The SEM-EDX result is redrawn at its bottom. By accident, the PnM-Nb sample forms the 2D square lattice networked by oxygen, whose configuration is the same as that of the copper--oxygen plane, the essential component of high-$T_{c}$ cuprate superconductors. Of course, there are several differences between them: the lattice constant $a$, the order of nm or 20 $\si{\micro m}$, and the composite metal, Cu or Nb. Yet the exact mechanism behind cuprate superconductors has not been fully elucidated. At the end of this paper, a link between the ``crystal'' structure and superconductivity is suggested courageously.

\subsection{Methods}
\label{sec:met}
First, temperature ($T$) dependence of the electrical resistance ($R$) was investigated. The \emph{R--T} measurement was performed using the PPMS (Quantum Design) under zero magnetic field. The sample chips were mounted on a PPMS sample puck using vacuum grease (Apiezon N, M\&I Materials), and the electrical contacts for the sample to the PPMS sample puck were made by aluminum wire bonding. The resistance was measured by the two-probe method using the electrical pads denoted in \cref{fig2}; pads number 1 and 2 were used. The PPMS was operated in the AC drive mode with the standard calibration mode, and the number of readings taken was twenty five. That is, at each temperature, positively and negatively oscillating 8.33-Hz square-wave excitation current with the amplitude of $\pm$10 $\si{\micro A}$ was repeatedly applied to the sample twenty five times, and the output voltage was obtained by averaging output values to minimize the DC offset error; all of these procedures were done by the PPMS automatically. One \emph{R--T} cycle consists of a cooling process from 300 K to 2 K and a subsequent warming process from 2 K to 300 K. The \emph{R--T} cycle (300 K $\rightarrow$ 2 K $\rightarrow$ 300 K) was continuously repeated for eight times.

Second, after finishing the \emph{R--T} measurement, the sample chip was taken out of the PPMS and was installed in the MPMS (Quantum Design), and its temperature ($T$) dependence of the magnetization ($M$) was investigated. The sample chip was positioned in a plastic straw. After the temperature was lowered down to 4.2 K under zero magnetic field, the magnetic field of 1000 Oe was applied, perpendicularly to the sample surface, and then the centering procedure was performed. The magnetization $M$ of the sample chip was scanned by moving the entire straw through the SQUID ring. The oscillation amplitude, frequency, cycles to average was 0.3 cm, 4 Hz, 40 cycles, respectively. The number of scans per measurement was three. While warming the sample chip from 4.2 K to 300 K, the zero field cooling (ZFC) measurement was performed. After that, the field cooling (FC) measurement was subsequently performed while cooling the sample chip from 300 K to 5.1 K with the applied magnetic field of 1000 Oe unchanged.{\if0 \emph{M--T} characteristics of an as-fabricated PnM-Nb sample, to which the above \emph{R--T} cycle had not yet been applied, were also measured using the same MPMS protocol.\fi}

236 days later, the above \emph{M--T} cycles were applied to the same sample again, and the temperature was raised to 300 K with the applied magnetic field of 1000 Oe unchanged. Subsequently, the applied field was once decreased to 0 Oe, and then applied magnetic field ($H$) dependence of the magnetization ($M$) was investigated at 300 K. The oscillation amplitude, frequency, cycles to average, number of scans was 0.5 cm, 4 Hz, 40 cycles, three, respectively.

Also, the critical magnetic field ($H_{c}$) at 300 K was investigated using the PPMS. The sample used for this measurement was another PnM-Nb which preserved zero resistance at 300 K during an \emph{R--T} cycle performed in advance. Under various magnitudes of perpendicular magnetic field, current was applied to the sample, and the output voltage was measured by the four-probe method using the electrical pads denoted in \cref{fig2} (pads number 1, 2 for applying current; 3, 4 for measuring voltage). The delta mode (6221/2182A combination, Keithley), which can minimize constant thermoelectric offsets, was externally connected to the sample in the PPMS. The current pulse with a width of 10 ms and a period of 100 ms was applied, and the output voltage was measured in the minimum range of 10 mV.

\section{Results}
\label{sec:res}
\subsection{Resistance Drop}
\label{sec:rdrop}
\begin{figure}[h!]
\centering
\includegraphics[width=0.89\linewidth]{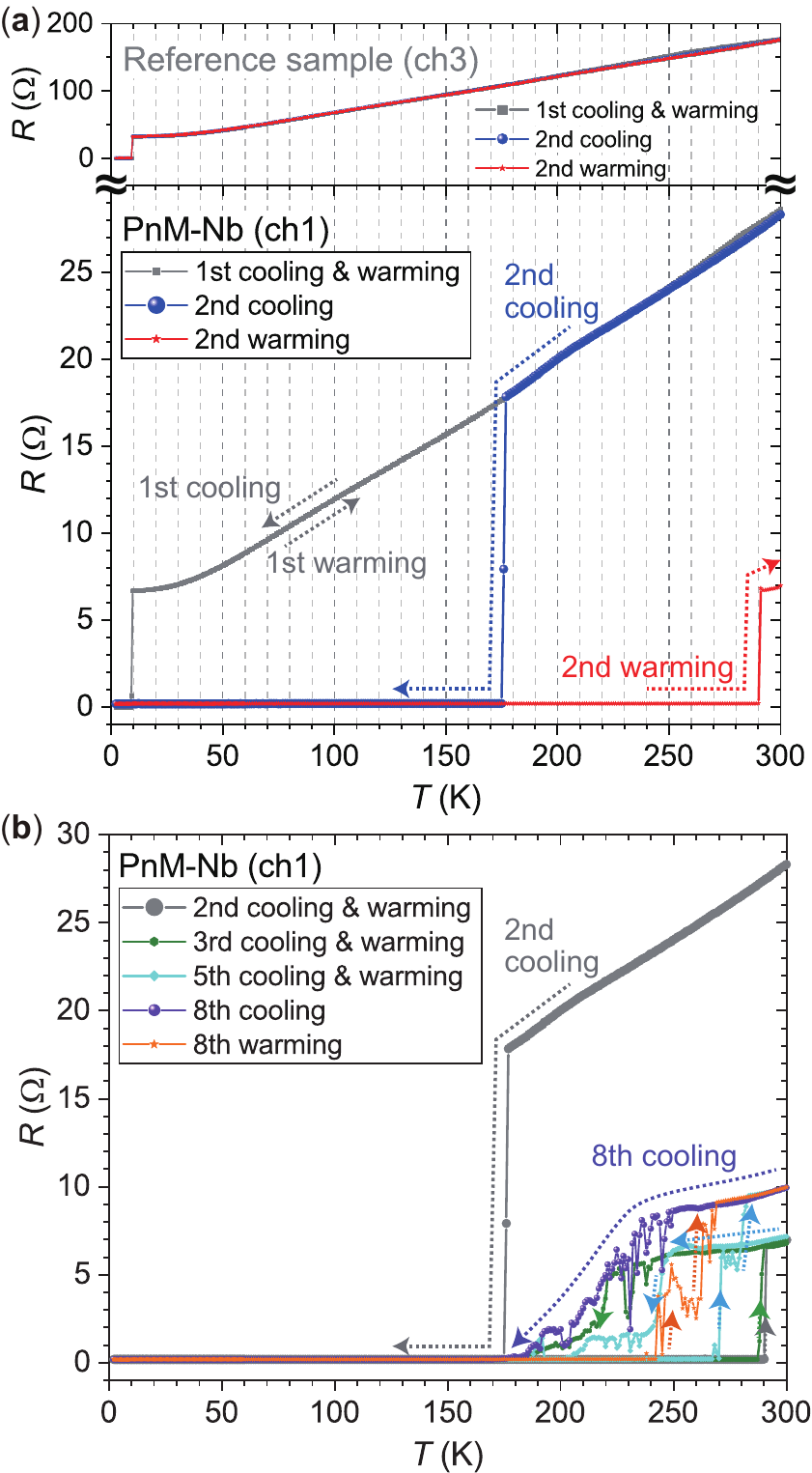}
\caption{\textbf{Resistance drop.} (\textbf{a}) Electrical resistance versus temperature of a reference Nb sample (upper panel) and that of the PnM-Nb sample (bottom panel). They were measured at the same time in the PPMS under the same condition. Only \emph{R--T} results for the first and second cycles are shown. (\textbf{b}) \emph{R--T} results of the PnM-Nb sample for the rest of temperature cycles. The whole raw data can be found at Zenodo~\cite{Zenodo2022}.
\label{fig4}}
\end{figure}

\autoref{fig4}a shows the temperature ($T$) dependence of electrical resistance ($R$) of the PnM-Nb sample under zero magnetic field. The upper panel shows that of a reference Nb sample, which was mounted on the PPMS sample puck together with the PnM-Nb sample and was measured at the same time under the same condition. (The PPMS can measure three samples at the same time.) As shown, the reference sample undergoes the superconducting (SC) transition normally at the usual $T_{c}$ for Nb both for the first and second temperature cycles. Although there shows only the first and second cycles, the result is the same for all the rest of the eight \emph{R--T} cycles; it can be confirmed in the raw data deposited at Zenodo~\cite{Zenodo2022}. The standard deviation of the onset SC transition temperature for all the eight \emph{R--T} cycles is 9.0000$\pm$0.0006 K, indicating that during the whole eight \emph{R--T} cycles the thermometer of the PPMS was working properly and that the temperature profiles of \emph{R--T} results are accurate. That is, whatever anomaly the PnM-Nb sample exhibits any criticism arising from thermometry is invalid.

In the first cooling process (gray curve in \cref{fig4}a), the PnM-Nb sample undergoes an SC transition at 9 K, which is the usual $T_{c}$ for Nb, and returns to the normal state at the same $T_{c}$ in the subsequent warming process. During the first temperature cycle, the PnM-Nb sample thus exhibits usual SC properties of the well-known conventional superconductor Nb, the element that constitutes the sample. In subsequent temperature cycles, by contrast, the sample begins to exhibit drastic changes. In the second cooling process (blue curve), the $R$ suddenly drops at 175 K, and the sample keeps the zero-resistance state down to 2 K. Due to the two-probe method adopted for this measurement, a tiny resistance residing in electrical pads ($\lesssim 0.2~\si{\ohm}$) remains in the $R$. The residual resistance aside, the sample itself shows absolute zero resistance. In the subsequent warming process (red curve), the zero-resistance state keeps going through 175 K, and a finite resistance appears at 290 K. The 2D PnM-Nb sample thus exhibited high $T_{c}$'s.

A possible reason for the warming curve that deviates from the cooling one and for the necessity of repeating temperature cycles to obtain the zero-resistance state is discussed later with the help of an already forgotten superconducting principle, \emph{giant atom}, and its comprehensive theory.

Uninterruptedly, \emph{R--T} measurements were repeatedly performed. The result is shown in \cref{fig4}b. The resistance drop, which was very sharp in the second temperature cycle, is obviously broadened by increasing the number of temperature cycles. Also, $R$ at 300 K does not return to the initial value of 28 $\si{\ohm}$, implying that the sample after the \emph{R--T} cycles is no longer a normal metal. The broadening of resistive transition is often observed when magnetic vortices are involved in the SC transition, being typical for the so-called type-II superconductors. The more vortices in the material, the wider the resistive transition becomes. Therefore, if this is a true study of superconductivity, the experimental fact of the resistive broadening indicates that the sample is intrinsically type-I but is changed into type-II where vortices are activated, by repeating \emph{R--T} cycles. Some might assume that type-I or type-II is material dependent and hence the change cannot be made in an identical material. But it is not true. Whether type-I or type-II originates from the Ginzburg-Landau theory which is postulated as a `phenomenological' model, meaning that whether type-I or type-II does not depend on what the material is. Another type just indicates another phenomenon is happening in a material. Of course, such a type change in a sole material has not been reported to date, but it is also true that such a sample prepared by physical microfabrication as the one in this study has not been tested before.

All questions above regarding vortices are answered in the following section by considering magnetic properties of the sample. Before beginning, it is noteworthy from \cref{fig2} that in the PnM sample there are a lot of empty spaces that would be suitable for vortices to reside---the array of voids. The orderly arrayed voids are qualitatively different from defects that are randomly distributed in a material. Such voids will give distinct consequences.

\subsection{Magnetic Flux Expulsion, i.e., the Meissner Effect}
\label{sec:meiss}
\begin{figure}
\centering
\includegraphics[width=1.0\linewidth]{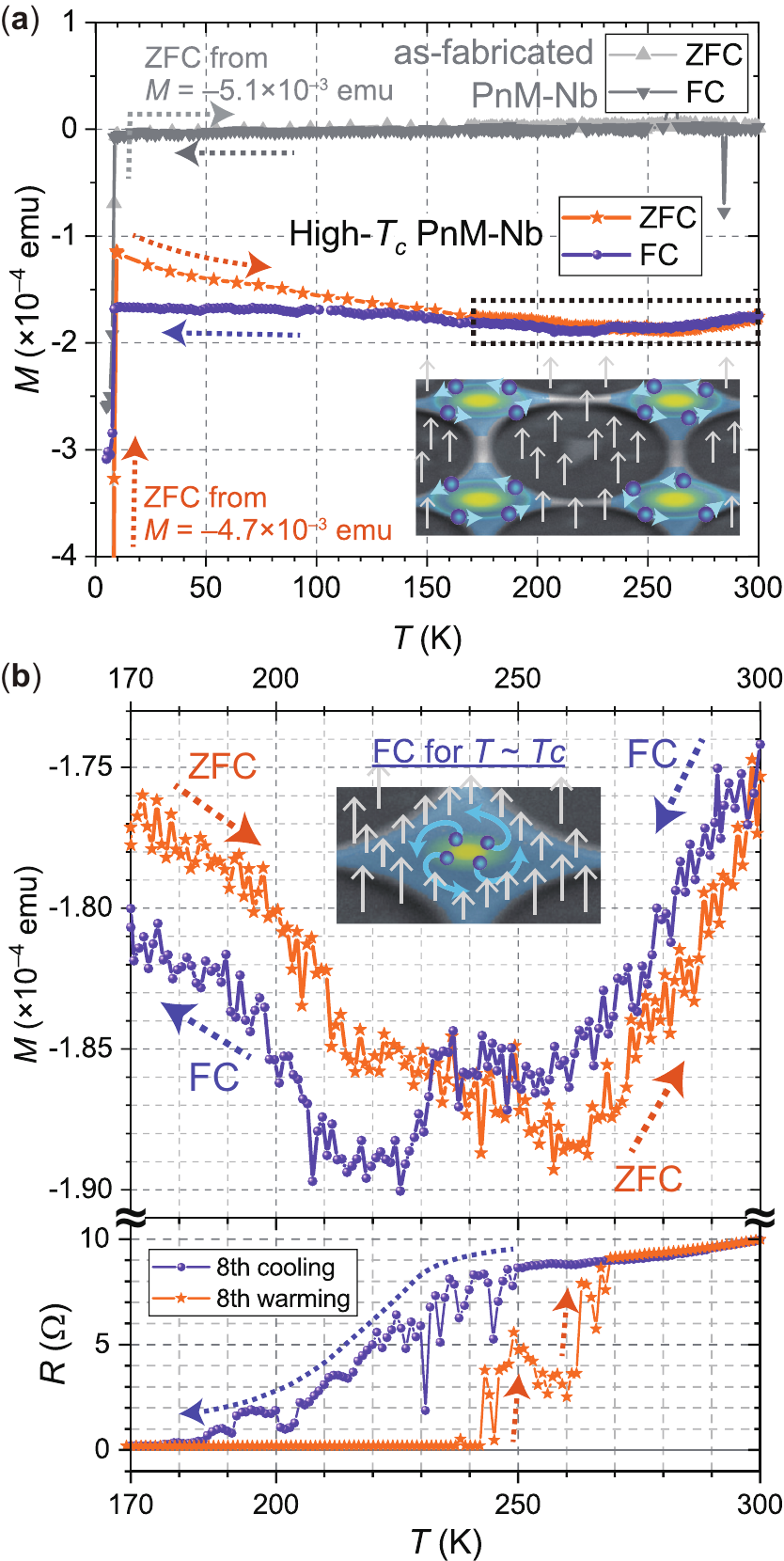}
\caption{\textbf{Magnetic flux expulsion.} (\textbf{a}) Magnetization versus temperature of an as-fabricated PnM-Nb sample (gray colors) and that of the PnM-Nb sample that exhibited high $T_{c}$'s in the preceding \emph{R--T} measurement (orange and purple colors). Applied magnetic field for both the \emph{M--T} measurements was 1000 Oe, perpendicular to the sample surface. (Inset) Schematic illustration of the PnM-Nb sample during the ZFC process for $T<T_{c}$ using a false-color SEM. Out-of-plane gray arrows, applied magnetic flux. In-plane cyan arrows, circulating negative charge particles in the counterclockwise direction in such a way as to shield the interior of superconducting ``islands'' from the applied magnetic flux. (\textbf{b}) Enlarged \emph{M--T} result in the temperature range of 170--300 K, surrounded by the black dotted square in (a). (Inset) Schematic illustration of the magnetic flux expulsion during the FC process, the Meissner effect. (Bottom panel) Duplicated \emph{R--T} result of the eighth cycle in \cref{fig4}b, which was performed just before this \emph{M--T} measurement.
\label{fig5}}
\end{figure}

The gray curves in \cref{fig5}a are temperature ($T$) dependence of magnetization ($M$) of an as-fabricated PnM-Nb sample, to which the \emph{R--T} procedure had not yet been applied and therefore that did not have a high $T_{c}$ yet. As the temperature was raised from 4.2 K in the ZFC process, the $M$ increased from $-5.1\times 10^{-3}$ emu to zero at $T\approx$ 9 K as usual, because superconducting (SC) diamagnetism of Nb constituting the sample disappears above 9 K. The as-fabricated sample kept the approximate zero value up to 300 K, and its subsequent FC curve down to 9 K roughly retraced the ZFC one as usual. The approximate zero value above 9 K indicates that both diamagnetism of the Si substrate supporting the microfabricated Nb film and paramagnetism of the entire Nb film including the microfabricated PnM-Nb sample part (see \cref{fig2}) are negligibly small. Below 9 K on the other hand the FC curve deviated from the ZFC one because applied magnetic flux are trapped in the Nb film when it undergoes the SC transition at 9 K, thus exhibiting a smaller magnitude of SC diamagnetism ($M\approx -2.6\times 10^{-4}$ emu) than that of the ZFC one, being typical for the type-II superconductor Nb.

The orange curve in \cref{fig5}a is the ZFC result of the PnM-Nb sample having anomalous high $T_{c}$'s after the \emph{R--T} cycles (\cref{fig4}b). As the temperature is raised from 4.2 K, the $M$ increases from $-4.7\times 10^{-3}$ emu at $T\approx$ 9 K, because SC diamagnetism residing in the Nb film disappears at 9 K. Note the magnitude of the ZFC value below 9 K is almost the same as that of the as-fabricated sample, indicating that the tilted angle of the sample chip in the MPMS plastic straw was almost the same for both the independently performed \emph{M--T} measurements. (The tilted angle affects the demagnetizing factor, which affects the magnitude of $M$.) Therefore, it is reasonable to compare the independently obtained two \emph{M--T} results.

In contrast to the $M$ of the as-fabricated sample that reached zero at 9 K (gray), the $M$ of the anomalous high-$T_{c}$ sample does not reach zero at 9 K (orange), remaining negative. As mentioned above, neither diamagnetic Si nor paramagnetic Nb affects the value of $M$. Only the microfabricated PnM sample part having anomalous high $T_{c}$'s can take responsibility for the anomaly.

For $T>$ 9 K, the ZFC value (orange curve in \cref{fig5}a) gradually decreases from $M\approx -1.1\times 10^{-4}$ emu to more negative values as $T$ is raised. When considering the \emph{R--T} result, the sample in this temperature range is in the superconducting (SC) state. Therefore, the \emph{M--T} curve should show a fundamentally flat temperature dependence if there were no magnetization other than the SC diamagnetism. A possible reason for the discrepancy is flux trapping. As shown in a false-color SEM in the inset, there is a non-material part in the PnM sample---void. Because of its large diameter of approximately 20 $\si{\micro m}$, applied flux easily invades the void. Once it invades, it remains trapped and moves together with the sample. That is, the SQUID ring detects an extra magnetization in addition to the SC diamagnetism residing in the material part. Since the direction of the extra magnetization is parallel to the applied field, the value of the measured $M$ increases, concealing the flatness of the temperature independent SC diamagnetism.

However, the extra magnetization owing to flux trapping is unfavorable to thermodynamic equilibration. As shown soon, the critical field $H_{c}$ for the PnM sample is very large. Because of its significantly large $H_{c}$, the thermodynamic equilibrium state of this sample under the field of 1000 Oe during this \emph{M--T} measurement is not the intermediate state but the perfect shielding state. For such a superconducting sample, the extra magnetization residing in the void is nothing but an unwanted source of thermodynamic nonequilibration. In other words, the extra magnetization owing to flux trapping decreases its magnitude as the temperature is raised, and the ZFC curve exhibits thus monotonically decreasing behavior.

As the temperature is raised further to 300 K, by contrast, the ZFC curve stops decreasing and alters its trend upward. The \emph{M--T} result in the temperature range of 170--300 K is enlarged on the upper panel of \cref{fig5}b. The lower panel shows the eighth \emph{R--T} result in the same temperature range, duplicated from \cref{fig4}b, which was performed just before this \emph{M--T} measurement. Precisely describing the complicated dynamics is difficult at this time. Yet it is remarkable that the $M$ measured in the ZFC `warming' process (orange) is flipping its trend upward in the temperature range of 250--270 K, which is the same temperature range where the resistance $R$ started to rise in the eighth \emph{R--T} `warming' process (orange). Based on the assumption that the appearance of $R$ is due to the disappearance of superconductivity, it must be reasonable to conclude that the flip of $M$ across 250--270 K is the indication that the SC diamagnetism residing in the PnM material part is disappearing or at least weakening.

Uninterruptedly, the \emph{M--T} measurement was continued while `cooling' the sample from 300 K, that is, the field cooling (FC) measurement was performed. As the temperature is lowered, the FC value (purple) decreases. The lowering of $M$ indicates that the applied magnetic flux is expelled from the interior of an examined specimen, and it cannot be accounted for by any physics other than superconductivity, the Meissner effect. The FC value reaches its minimum at the temperature below 230 K, which is roughly consistent with the temperature range where the resistance $R$ started to vanish in the eighth \emph{R--T} `cooling' process (purple). When going superconducting, electrons start the azimuthal motion (see inset of \cref{fig5}b), expanding the interior superconducting domain radially outward. This is the Meissner effect. Hence it is not a trivial question which arrives first, the resistance drop or the magnetic flux expulsion. Either way, the occurrences of resistance drop and magnetic flux expulsion in the same sample in the same temperature range are the indication of the occurrence of superconductivity.

The FC curve (purple) flips its trend upward below 210 K, which must be attributed to flux trapping in a non-material part, i.e. void, as mentioned earlier. The flux trapping is going to happen after the PnM material part undergoes the superconducting transition. As the direction of the trapped flux is parallel to the applied field, the value of $M$ increases. Below 170 K, however, the FC curve exhibits a relatively flat temperature dependence in contrast to the ZFC curve (see \cref{fig5}a). For the ZFC measurement, the sample was already cold before the field was applied and, of course, did not know whether the field would be applied or not. Therefore, there was no choice for the sample other than to admit the extra magnetization in the voids. For the FC measurement, by contrast, the sample was cooled under the existence of the applied field. That is, there was a chance for the emergent superconducting screening currents to draw the most suitable geometric pattern that nullifies not only the field invading into the interior of the material part but also the other one trying to reside in the neighboring voids, in order to achieve the perfect shielding state demanded by its own large $H_{c}$. Therefore, the thermodynamically unfavorable magnetization arising from the voids was weakened during the FC measurement to the extent possible, and the FC curve exhibits thus relatively flat temperature dependence.

Finally, it is noteworthy that the value of $M$ still remains negative at 300 K, approximately $-1.75\times 10^{-4}$ emu regardless of whether the ZFC or FC process. By taking into account the \emph{M--T} result of the as-fabricated sample (gray curves in \cref{fig5}a), there is nothing in the sample inherently being responsible for the large negative value. Additionally, the clear separation between the ZFC and FC curves (orange and purple) cannot be made by any kind of magnetic impurities. So there is no possibility other than supposing that the negative value at 300 K is attributed to the SC diamagnetism arising from the high-$T_{c}$ PnM-Nb. Since the minimum $M$, approximately $-1.89\times 10^{-4}$ emu, is reached at lower temperatures, it must be reasonable to assume that not whole but just some part of the PnM-Nb is in the SC state at 300 K. In other words, superconductivity and nonsuperconductivity are likely to coexist in the PnM-Nb at 300 K. If so, the appearance of resistance at high temperatures (bottom panel of \cref{fig5}b), where the resistance did not return to the initial metallic value of 28 $\si{\ohm}$, might be attributed to the interaction between superconductivity and nonsuperconductivity, for instance, the resistive vortex motion driven by applied current during the \emph{R--T} measurement. To evaluate quantitatively, however, an estimation on the Josephson plasma frequency for the PnM sample is necessary and is beyond the scope of this study. This last remark and a condition under which the superconducting state at 300 K starts to melt will be shown in the forthcoming paper.

Anyway, there is a much more significant common feature in the \emph{R--T} and \emph{M--T} results (see \cref{fig5}b). That is, the zero resistance state survived at higher temperatures in the \emph{R--T} `warming' process than that in the cooling one, and the minimum diamagnetic value was achieved at higher temperatures in the ZFC `warming' process than that in the FC cooling one. These experimental facts indicate the benefit of `warming' to superconductivity. This will be discussed later in this paper to be connected with the lattice expansion which is likely to happen especially in the warming process. The lattice expansion seems very compatible with the self-standing PnM sample that has a degree of flexibility in the way the lattice is expanded (\cref{fig3}a).

\subsection{Flux Trapping in the Array of Voids}
\label{sec:rem}
Without flux trapping, neither the separation between the ZFC and FC curves (\cref{fig5}a) nor the rise in $M$ for the temperature lower than $T_{c}$ (\cref{fig5}b) could have been explained. Furthermore, the value of $M$ remaining negative at 300 K was interpreted as indicating a remnant of superconductivity at the temperature. If really applied flux was trapped in the voids during the \emph{M--T} measurement and if really some part of the PnM-Nb is still in the SC state at 300 K, then the trapped flux should be detected still at 300 K.

\begin{figure}[h!]
\centering
\includegraphics[width=0.95\linewidth]{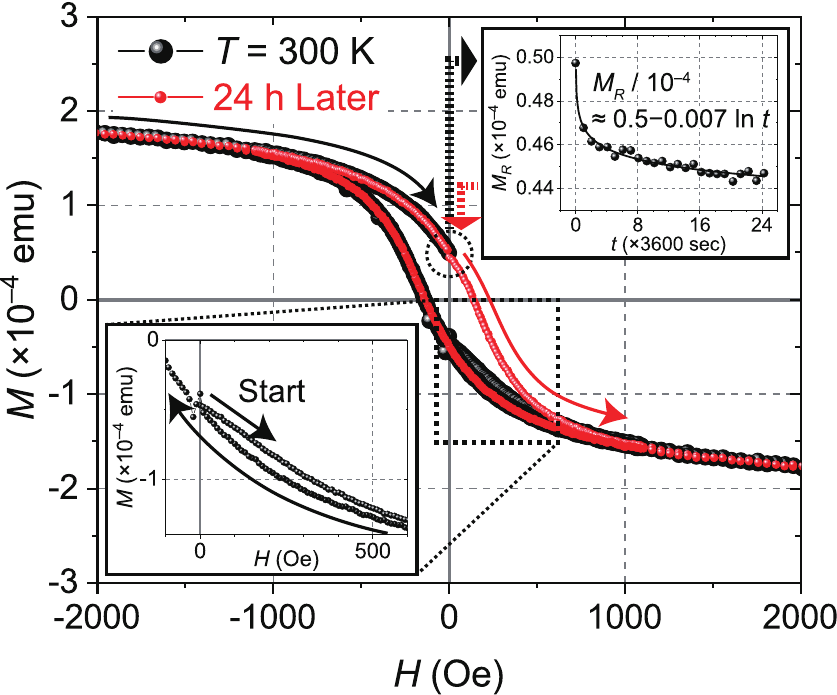}
\caption{\textbf{Flux trapping at 300 K.} Magnetic field dependence of magnetization of the PnM-Nb sample (black color) and that measured again 24 hours later (red color).{\if0 Lower-left inset, enlarged \emph{M--H} result surrounded by the black dotted square in the main panel. Only the result for the first cycle is shown. Upper-right inset, time evolution of the remnant magnetization at $H=$ 0 Oe.\fi} Both were measured at 300 K.{\if0 For details, read text.\fi}
\label{fig6}}
\end{figure}

\autoref{fig6} shows applied magnetic field ($H$) dependence of the magnetization ($M$) of the PnM-Nb sample, measured at 300 K. As shown in the lower-left inset, the \emph{M--H} measurement was started with the applied field increased from 0 Oe. It is remarkable that at $H=$ 0 Oe the $M$ already shows a non-zero value, approximately $-0.5\times 10^{-4}$ emu. After the FC measurement performed in advance of this \emph{M--H} measurement, the temperature was raised to 300 K with the applied field of 1000 Oe unchanged, then, after the temperature reached 300 K, the applied field was decreased to 0 Oe{\if0, and this \emph{M--H} measurement was subsequently started\fi}. Therefore, if really some part of the PnM-Nb was still in the SC state at 300 K, the decreasing magnetic field applied at 300 K should reduce the magnitude of trapped flux in voids, and, in turn, supercurrents in the PnM-Nb material part should circulate around the voids in such an inductive way as to prevent the trapped flux from decreasing its magnitude. In other words, in a decreasing field a supercurrent circulating around a void increases its magnitude. By taking into account the fact that such a circulating supercurrent consists of negative charge particles and by taking a look at the schematic illustration in the inset of \cref{fig5}a, it turns out that the tangential direction of such circulating negative charge particles around the void is the same as that of negative charge particles circulating around the periphery of superconducting ``islands'' generating diamagnetism. That is, the SC diamagnetism is enhanced in a decreasing field, therefore, even when the field is decreased to zero, the value of $M$ still remains negative. That's why at the beginning of this \emph{M--H} measurement the $M$ already showed the negative value instead of zero.

As the virgin curve shows (lower-left inset of \cref{fig6}), the SC diamagnetism increases with the increasing $H$. The `low'-field diamagnetism is a usual characteristic of superconductivity, but the monotonic behavior is lasting even at $H=$ 2000 Oe, a maximum field strength applied for this \emph{M--H} measurement. As shown soon, the critical field $H_{c}$ for the PnM-Nb is very large, and the 2000 Oe turns out to be indeed `low' enough when compared to the $H_{c}$.

When $H$ is subsequently decreased from 2000 Oe, the SC diamagnetism decreases, thus $M$ increases. It is remarkable that the hysteresis behavior begins at 1000 Oe, which is the same magnitude of the magnetic filed applied to the PnM sample during the preceding \emph{M--T} measurement{\if0 and hence is the one that is supposed to remain trapped in the array of voids\fi}. The decreasing $H$ tries to reduce the magnitude of trapped flux in the array of voids. Due to the same physics explained above, in the decreasing field the SC diamagnetism is enhanced, thus the measured $M$ becomes lower when $H$ is decreased versus when it is increased as can be seen in the lower-left inset. Upon lowering of $H$ from a positive to negative value the value of $M$ fluctuates a little bit. The central peak at $H=$ 0 Oe is another usual characteristic of superconductivity in addition to the low-field diamagnetism~\cite{SciAdv2020}.

When $H$ is increased from $-2000$ Oe and reaches 0 Oe (main panel of \cref{fig6}), the PnM-Nb exhibits a positive $M$, approximately $0.5\times 10^{-4}$ emu. This is a remnant magnetization. At 300 K and in the absence of applied field, the time evolution of the remnant magnetization $M_{R}$ was measured for 24 hours. The result is shown in the upper-right inset. At finite temperatures, thermal energy may allow flux lines to jump from one void to another in response to flux-density gradient, hence there may be an observable decrease of the magnitude of trapped flux with time~\cite{TinkhamBook}. Since any creep will relieve the gradient, the creep gets slower and slower, hence the time dependence is logarithmic as indicated by the experimental result. Given that the driving force is proportional to the magnitude of trapped flux, the exponential dependence of the creep rate on the driving force is
\begin{equation}
\frac{dM_{R}}{dt}\propto -e^{M_{R}/C_{1}},\label{eq:Mr}
\end{equation}
where $t$ is time, which has the solution $M_{R}=C_{2}-C_{1}\ln t$ where the constants $C_{1}$ and $C_{2}$ can be obtained from the fitting curve shown in the figure. Then it is possible to estimate how long the trapped flux persists in the array of voids and hence how long the circulating supercurrents will take to die out at 300 K. It will take $t=e^{C_{2}/C_{1}}\approx e^{0.5/0.007}\approx 10^{31}$ sec $\approx 10^{23}$ years.

When compared to the pioneering study for a tubular sample made of hard-superconductor NbZr measured at 4.2 K~\cite{Kim1962} or to the recent study for a granular superconductor made of graphite powder measured at 300 K~\cite{Esquin2012}, the lifetime is $10^{69}$ or $10^{370}$ times shorter. Yet, when considering the thermodynamically unstable nature of trapped flux in the PnM-Nb having a large $H_{c}$ and hence preferring to remain the perfect shielding state, the lifetime of the unfavorable flux trapping is still long, long enough in any practical sense. Because the PnM sample is apparently neither hard nor granular, the Bean model is of course not applicable to this study.

After the \emph{$M_{R}$--t} measurement, i.e. 24 hours later, the \emph{M--H} measurement was performed again with $H$ increased from 0 Oe (red curve in \cref{fig6}). As shown, it retraces the initial hysteresis loop, indicating that the PnM-Nb is indeed trapping magnetic flux and hence can sustain persistent currents without dissipation. The remnant of 300-K superconductivity has thus proven to be real.

\subsection{Critical Field}
\label{sec:cri}
\begin{figure}[h!]
\centering
\includegraphics[width=0.95\linewidth]{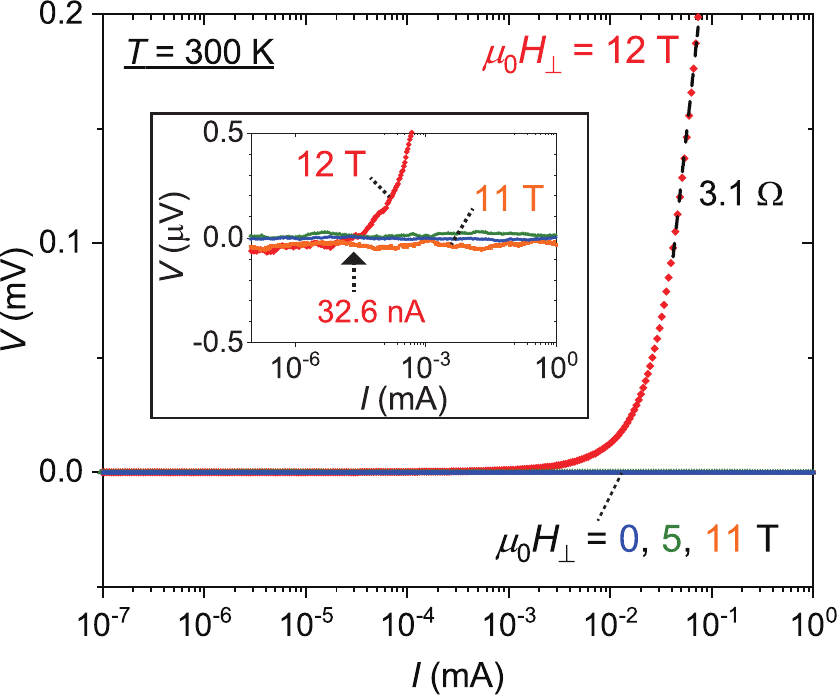}
\caption{\textbf{Critical field at 300 K.} Output voltage versus applied current under various magnitudes of perpendicular magnetic field, measured at 300 K. Inset, zoomed-in plot.
\label{fig7}}
\end{figure}

As mentioned previously, the critical field $H_{c}$ for the PnM-Nb was investigated at 300 K. Under various perpendicular magnetic fields of $\mu_{0}H_{\perp}=$ 0, 5, 11 and 12 T, the output voltage was measured while applying current. The result is shown in \cref{fig7}. For $\mu_{0}H_{\perp}\leq$ 11 T, the voltage does not increase for applied currents lower than 1 mA. At $\mu_{0}H_{\perp}=$ 12 T, by contrast, a finite voltage is observed for current exceeding a small value of 32.6 nA as shown in the inset. Hence, the intrinsic $\mu_{0}H_{c}$ at 300 K for the 2D PnM-Nb without applied current shall be a little bit larger than 12 T.

By taking into account the initial sharp resistance drop (gray curve in \cref{fig4}b), the PnM-Nb is intrinsically the so-called type-I superconductor, and therefore its $H_{c}$ corresponds to the thermodynamic critical field. As explained previously, the large thermodynamic critical field was responsible for the clear separation between the ZFC and FC curves (\cref{fig5}a). Even if an unknown demagnetizing factor of the PnM sample yields a 100 times larger magnitude than that of the actual field applied during the \emph{M--T} measurement (1000 Oe), the product is still lower than the $H_{c}$. That is, the perfect shielding state is indeed favorable for the PnM-Nb sample during the \emph{M--T} measurement. In other words, flux trapping in the array of voids is thermodynamically unfavorable for the PnM-Nb sample. That's why the ZFC curve exhibited the decreasing behavior as the temperature was raised. Also, whenever possible the flux trapping is suppressed as much as possible, that's why the FC cooling curve exhibited the relatively flat temperature dependence. Nevertheless, once trapped under isothermal condition, it remains trapped for a sufficiently long time as shown in the previous section.

Anyway, the value of $\mu_{0}H_{c}\gtrsim$ 12 T is much too large when considering the fact that the largest known critical fields for standard type-I superconductors are of order 0.05 T. Indeed there is a huge difference, but it turns out that the absurdly large $H_{c}$ is a natural consequence of 2D superconductivity. This is discussed in the following section on the basis of a superconducting principle, \emph{giant atom}, which was already forgotten a long time ago.

\section{Discussion}
\label{sec:disc}
\subsection{The $H_{c}$ and \emph{Giant Atom}}
\label{sec:giant}
In 1933, the experiment of Walther Meissner and Robert Ochsenfeld revealed that an externally applied field was expelled from the interior of a supraconductor. The discovery, now known as the Meissner effect, surprised physicists in that era, since the general solution of Maxwell's equations predicted that the original field had to persist for ever in the supraconductor. However the argument over the frozen-in magnetic fields in supraconducting bodies was immediately resolved by Fritz and Heinz London brothers in 1934--35~\cite{LondonBro1934}. They showed the existence of screening current flowing around the superconductor in such a way as to shield the interior from the applied magnetic field. The endlessly circulating screening current was further considered by F. London, being very strongly reminded of Gordon's formulae for electric current and charge in his relativistic formulation of Schr\"{o}dinger's theory. That is, F. London considered that even \emph{one single electronic state} was sufficient for representing the electromagnetic behavior of a superconductor with all its various possible currents. He thus coined the phrase ``\emph{macroscopic quantum phenomenon}'' and characterized the electromagnetic behavior of a superconductor as ``\emph{being the same as that of a single big diamagnetic atom}''~\cite{London1937}. In this line of thought, he predicted the existence of the flux quantum $\phi_{0}$, which was experimentally observed in 1961, seven years after his death in 1954.

Following the logical argument raised by F. London in 1937~\cite{London1937}, John Slater immediately considered a magnetic susceptibility of an atom, and, in order for the atom to have perfect diamagnetism, he gave the minimum diameter of the atom,
\begin{equation}
R_{S}=137\times 2a_{0}~(\approx \mathrm{14.5~nm}),\label{eq:Rs}
\end{equation}
where $a_{0}$ is the Bohr radius (0.529 \AA) and the number 137 comes from the fine structure constant ($\alpha\equiv \frac{e^{2}}{\hbar c}\approx \frac{1}{137}$)~\cite{Slater1937}. The Slater's atom is the basic one of what F. London called ``\emph{a single big diamagnetic atom}'', a.k.a., the ``\emph{giant atom}''~\cite{LondonBro2011}.

They are teaching us that a wavefunction around the core of the \emph{giant atom} is stationary, forming a loop, the same as that of the usual Bohr's atom, but is radially expanded, taking only the quantized angular momentum corresponding to $n\phi_{0}$ ($n=$ 1, 2, ...). This is also true for an atom that provides its wavefunction to a screening current under an applied magnetic field. For this reason, relatively large spheres are illustrated in the inset of \cref{fig5}a. (Four spheres are shown in a screening current circulating around the periphery of a superconducting ``island''.) In the screening current, each wavefunction has the expanded diameter $R_{S}$, carrying a single diamagnetic $\phi_{0}$ that resists the applied field. When does the applied field invade the interior of the superconducting island? It is when the screening current is filled with a bunch of Slater's atoms overlapping each other. Then the critical field can be calculated as
\begin{equation}
\mu_{0}H_{c}=\frac{\rm{\phi_{0}}}{\pi\times (R_{S}/2)^{2}}~(\approx \mathrm{12.5~T}).\label{eq:hc}
\end{equation}
This is what the experimental \cref{fig7} is indicating.

Such an absurdly large $H_{c}$ is thus verified by this simple equation. There is no doubt that the unrealistic prediction value of the equation was the primary reason why the \emph{giant atom} was abandoned by the superconductive community once upon a time. But this time the absurd prediction turns out to be true. Hence we have to think about the \emph{giant atom} again. Note this equation is valid only in a 2D perspective, indicating that 2D superconductivity is responsible for the experimentally confirmed large $H_{c}$. This consequence may reflect a key role of the 2D electron system that this study at the beginning expected of the 2D PnM.

\begin{figure}[h!]
\centering
\includegraphics[width=0.85\linewidth]{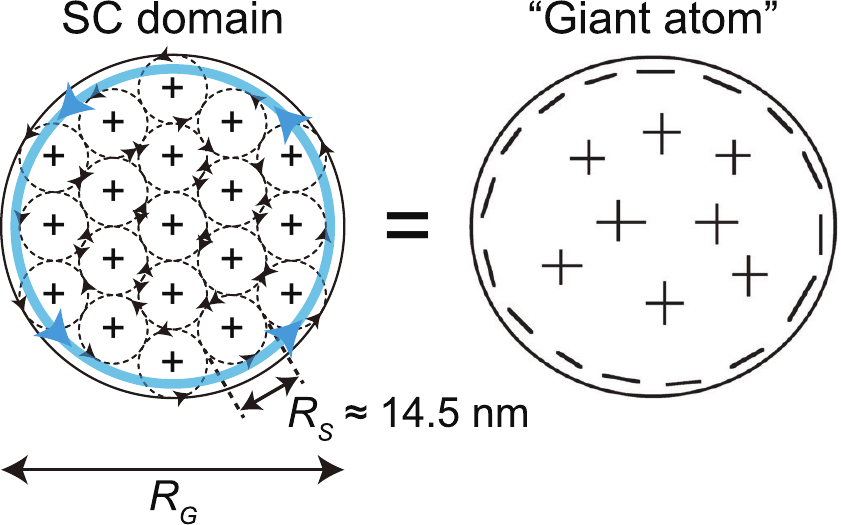}
\caption{Schematic superconducting (SC) domain with a diameter $R_{G}$, filled with Slater's atoms with the diameter $R_{S}$. Right panel, its superpositioned equivalent. ``$+$'' and ``$-$'' indicates a positive and negative charge, respectively. Cyan arrow, a screening current. The right panel is reprinted from ref.~\cite{HirschJAP2021}, with the permission of AIP Publishing.
\label{fig8}}
\end{figure}

\autoref{fig8} (left panel) shows a schematic superconducting (SC) domain, in which Slater's atoms are arranged in such a way as to be associated with the 2D \cref{eq:hc}. In the inner domain, the superposition of electrons orbits nullifies themselves, leaving electrons orbits only within a thin layer adjacent to the periphery. Under an external magnetic filed, the remaining electrons orbits at the periphery draw a supercurrent path (cyan arrow) circulating around the SC domain in such a way as to shield the interior from the applied field. The right panel shows the equivalent of the SC domain, reprinted from ref.~\cite{HirschJAP2021}. By the superposition, only positive charges, i.e. holes, remain in the bulk, and negative charges are globally expelled from the bulk to the periphery. Indeed it looks like a `giant atom'. The heavily concentrated holes exert repulsive Coulomb force on their own lattice. That is, the lattice is expanding.

In the following section, the remaining questions, why warming is beneficial to superconductivity and why repeating temperature cycles is necessary to obtain this superconductivity, are explored to the extent possible, with the help of a comprehensive theory of the \emph{giant atom}---the theory of \emph{hole superconductivity}.

\subsection{Benefit of the Lattice Expansion}
\label{sec:expansion}
Though London characterized a superconductor as being the same as a single big diamagnetic atom, even he could not solve ``\emph{which kind of interaction}'' could be made responsible for the appearance of such separated diamagnetic states~\cite{London1937}.

During the golden age of BCS theory from 1957 to 80s, extensive efforts to search for superconductors with a higher $T_{c}$ were carried out, and many superconducting inter-metallic alloys and compounds were discovered, giving rise to the then-record highest $T_{c}$ of 23 K observed in Nb\textsubscript{3}Ge~\cite{Test1974,Gavaler1974}. It is noteworthy however that the achievement was not by the BCS theory but by Bernd Matthias's effort~\cite{Matt1955,Matt1965,Matt1971,Matt1973}, who for the first time publicly railed against the BCS theory to emphasize the fact that the theory is not useful in predicting where to find high-$T_{c}$ superconductors~\cite{Maple2015}. Matthias brought to light a significant alternative: ``\emph{the simultaneous occurrence of lattice instabilities and high $T_{c}$ superconductivity}''~\cite{Chu:BCS}. The lattice instability is a logical extension of the Matthias's empirical rule that teaches us the importance of the number of valence electrons outside the filled shell of an atom~\cite{Matt1955,Matt1973}. Matthias thus paid attention to the \emph{reality} of superconducting materials. BCS on the other hand did not take into account the \emph{real} atom structure, in fact failing to find \emph{real} high-$T_{c}$ superconductors~\cite{HirschJAP2021}. When a conduction band is almost full of electrons, the electrical conductivity is by holes. This consideration is consistent with the fact of superconducting materials, that the great majority of materials that go superconducting have positive Hall coefficients. This fact was pointed out for the first time in 1932~\cite{Kikoin1932} and was also discussed in papers by others~\cite{Born1948,Feynman1957,Chap1979}. Among them, the statement by Richard Feynman in ref.~\cite{Feynman1957} published in 1957, several months earlier than that of the BCS theory, is especially remarkable: ``\emph{... if Frohlich and Bardeen could solve their model exactly, they still would not find superconductivity, since it would still involve only negative carriers.}''

The London and Slater's \emph{giant atom} describes the static view of superconductivity, only. What London could not achieve and what he has been demanding is the theory that finally forms the \emph{giant atom} in its own framework. Considering the above \emph{realities} of superconductors, namely, holes and the Matthias's lattice instability, Jorge Hirsch has proposed an alternative in 1989, the theory of \emph{hole superconductivity}~\cite{Hirsch1989}. His theory is necessary to explain experimental results of this study: why the superconductivity in the warming process could survive at higher temperatures than that in the cooling one and why repeating temperature cycles was necessary to obtain this superconductivity.

Holes propagating through a material destabilize the lattice, that's why they are called `antibonding', and therefore the lattice ions cannot avoid repelling each other, that is, the lattice `expands'~\cite{HirschWeb}. When the lattice expands, the wavefunctions of orbital electrons in the vicinity of each lattice point also expand, and the expansion lowers the quantum kinetic energy of each orbital electron,
\begin{equation}
E_{kin}=\frac{\hbar^{2}}{2mr^{2}},\label{eq:ekin}
\end{equation}
where $\hbar$ is the reduced Planck constant, $m$ is the electron mass and $r$ is the radial extension of the wavefunction, hence leading to the global expulsion of electrons, ultimately, forming the \emph{giant atom}. It is remarkable that the theory has finally reached the same conclusion as that given by London and Slater. Though the minimum diameter of the \emph{giant atom} was already given by Slater~\cite{Slater1937}, the maximum limit by contrast was not given and even not known if it exists or not; the future work must be exciting either way.

According to London, ``\emph{the most stable state of any system is not a state of static equilibrium in the configuration of lowest potential energy. It is rather a kind of kinetic equilibrium ...}''~\cite{LondonBook}. \emph{Hole superconductivity} describes the microscopic dynamics of holes, electrons, lattice instability, all at once, with respect to the ``\emph{lowering of quantum kinetic energy}'' and succeeds in connecting the microscopic dynamics to the macroscopic superconducting behaviors without discontinuity in scale. London may be satisfied this time. One of the macroscopic behaviors, Meissner effect, was studied in this paper and was described in the inset of \cref{fig5}b. The explanation given for it was, to be honest, taken from the theory of \emph{hole superconductivity}. The magnetic field lines are being expelled together with electrons expanding their movements outward, and these negative charge wavefunctions themselves also expand their radii outward to generate the superconducting screening current. For more spectacular consequences, see refs.~\cite{HirschSpin2007,HirschBook}.

Thus, the expansion rules superconductivity, and any expansion may be beneficial. The warming process as well makes the sample lattice expand rather than shrink, that's why the superconductivity was enhanced for the warming process. The fact that the self-standing lattice contains anomalous regions where the atomic distance is increased by 0.1 \AA~(\cref{fig3}a) should not be flatly ignored. More or less, it has a degree of flexibility in the way to be expanded. The necessity of repeating \emph{R--T} cycles, on the other hand, must be related to the lattice instability caused by the propagation of holes. During the \emph{R--T} cycles, an excitation current was applied to the sample to measure the resistance as a matter of course. When the propagating charge carriers are holes, which was the case for this sample consisting of Nb having a positive Hall coefficient~\cite{Chap1979}, the propagation destabilizes the sample lattice, leading to the above Matthias's conviction. Based on this consideration, applying current is indispensable for changing an as-fabricated sample into the superconducting one. In fact, no matter how many times the \emph{M--T} cycles were applied to an as-fabricated sample, the sample without applied current did not undergo the high-$T_{c}$ superconducting transition. Hence the \emph{R--T} cycles have to be carried out first to achieve this superconductivity. To predict when and how the sample undergoes the high-$T_{c}$ transition, better insight into the lattice instability and better control over in-plane stress (\cref{fig3}a) are crucial.

\subsection{Reproducibility}
\label{sec:repro}
To date, 18 PnM-Nb samples with a rectangular shape were examined, and 7 samples exhibited zero resistance at high temperatures. So the yield rate is 39\%. All the superconductivity were enhanced in a warming process. Additionally the zero resistance was attainable regardless of the sample size and regardless of whether the resistance was measured by the two or four probe method. However, the $T_{c}$ varied with the sample, taking some typical values: 35 K, 50 K, 100 K, and 175 K. Additionally the number of \emph{R--T} cycles repeatedly applied to the sample in order to change it into the zero-resistance state varied with the sample. A possible reason for those varieties and an effective way to enhance the yield rate are discussed below.

\begin{figure}[h!]
\centering
\includegraphics[width=0.9\linewidth]{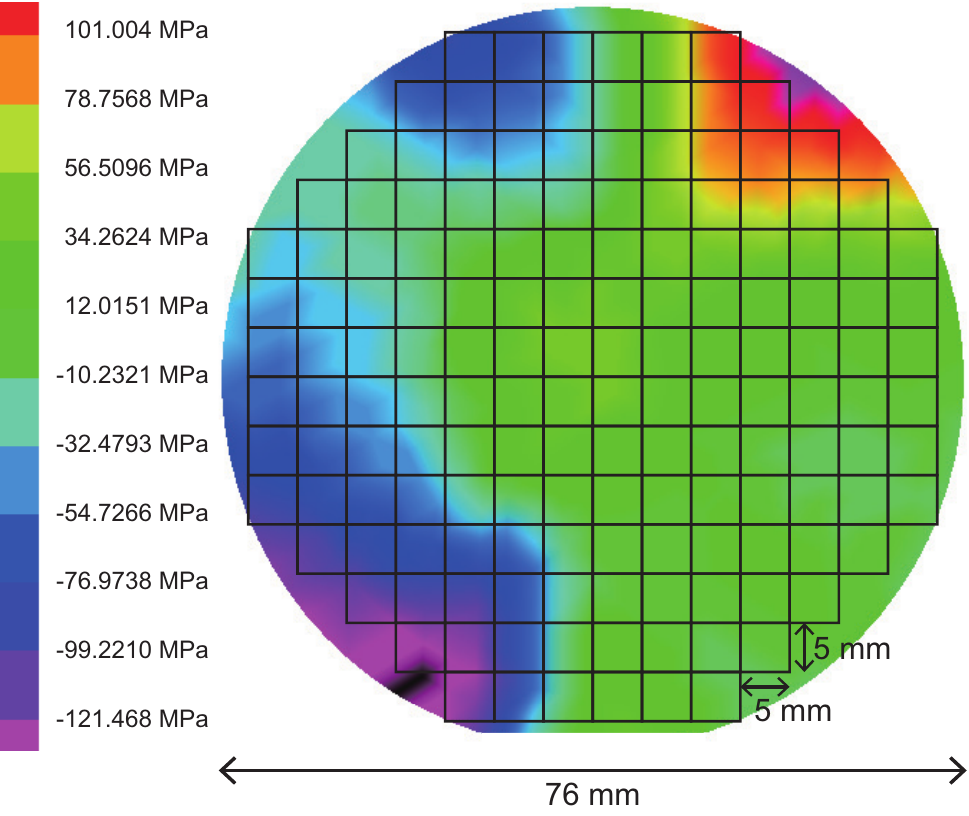}
\caption{In-plane stress of the Nb layer sputtered on the SiO\textsubscript{2}/Si wafer with the diameter of 76 mm. The size of a sample chip (5$\times$5 mm\textsuperscript{2}) is also indicated. The in-plane stress was measured using FSM 128NT (Frontier Semiconductor).
\label{fig9}}
\end{figure}

As described in \cref{sec:mat}, those samples were microfabricated on a 3-inch ($\approx$ 76-mm) wafer, and each sample chip with the size of 5-mm squares was taken from the wafer. As shown in \cref{fig9}, there is a distribution of in-plane stress in the Nb layer deposited on the SiO\textsubscript{2}/Si wafer. In-plane stress is widely distributed in the range from $-$120 to $+$100 MPa. That is, each sample chip has a different in-plane stress to each other, despite the used sputtering condition that was adjusted for zero in-plane stress. The different in-plane stress may yield a different magnitude of lattice expansion after the final HF dry etching process that removes the SiO\textsubscript{2} sacrificial layer under the Nb layer. And the different magnitude of lattice expansion (\cref{fig3}a) may also affect oxidization of the self-standing PnM sample (\cref{fig3}b). Therefore, if the high-$T_{c}$ superconductivity found in this study is truly promoted by the 2D oxygen network configuration, the different magnitudes of lattice expansion and oxidization may be highly important factors in determining whether the thus microfabricated PnM sample undergoes the superconducting transition or not and its $T_{c}$.

Hence, controlling in-plane stress and oxidization during the microfabrication process will be the next important step to increase the yield rate for the zero-resistance state. When taking into account the discussion in the previous section, a higher magnitude of lattice expansion seems preferable to the occurrence of superconductivity. Therefore, intentionally using another Nb sputtering condition that increases in-plane stress is worth considering. The higher in-plane stress will induce the higher lattice instability as well, which is also beneficial to the occurrence of superconductivity. This can be done by adjusting pressure and flow rate of Ar gas during the Nb sputtering deposition. Current oxidization, on the other hand, depends on natural oxidation. An alternative method, for example applying oxygen plasma ashing to the PnM sample prior to \emph{R--T} cycles, will be better to control oxidization. Investigating how they affect \emph{R--T} results will be the most significant study.

Since the zero-resistance state presented in this study was for the first time observed in March 2018, many \emph{R--T} measurements had been performed, but none of them had succeeded in reproducing the zero-resistance state. The above yield rate (39\%) is achieved after an inevitable condition during the \emph{R--T} cycles was found in November 2018, namely the slow cooling/warming rate{\if0 especially for the temperature range of 20--60 K in the first six temperature cycles\fi}. A typical \emph{R--T} recipe is summarized in \cref{tab1}. The necessity of the slow rate especially for the temperature range of 20--60 K may be related to oxygen already absorbed in the PnM sample, which is known to undergo an antiferromagnetic transition below 54 K and to change its solid structure ($\beta \leftrightarrows\gamma$ phase) at 43 K. If the oxygen transformation truly happens in the PnM, the Jahn-Teller theorem may have to be considered as another glue for the 2D phonon--electron interaction, besides phonon engineering. The central issue of the theorem is that ionic displacements and electronic motion cannot be decoupled~\cite{JT2022}. That is, the theorem is very compatible with the lowering of electronic kinetic energy \cref{eq:ekin} caused by the lattice expansion. This compatibility associated with the oxygen phase transformation has to be investigated together with the in-plane stress and oxidization discussed above.{\if0 Whether the Jahn-Teller theorem alone can be the underlying mechanism of high-$T_{c}$ cuprates depends on whether it accounts for the dynamics of Meissner effect~\cite{JT1996} and whether it forms the London and Slater's \emph{giant atom} in its own framework.\fi}
\begin{table}[h!]
\caption{\label{tab1}~Typical \emph{R--T} recipe$^*$ to make the zero-resistance state}
\vspace{-4mm}
\begin{center}
\footnotesize
\begin{tabular}{@{}llcc}
\toprule
\begin{tabular}{l}\emph{R--T} cycle\\number\end{tabular}&\begin{tabular}{c}Cooling--warming rate\end{tabular}&\begin{tabular}{c}Temperature\\approaching\\mode$^{\dag}$\end{tabular}&\begin{tabular}{c}Elapsed\\time\end{tabular}\\
\toprule
1st~--~6th&300~$\rightarrow$~~~60 K: 1.0 K/minute&S&\\
&~~60~$\rightarrow$~~~20 K: 0.5 K/minute&F&\\
&~~20~$\rightarrow$~~~~~2 K: 1.0 K/minute&S&\\
&~~~~2~$\rightarrow$~~~20 K: 1.0 K/minute&S&\\
&~~20~$\rightarrow$~~~60 K: 0.5 K/minute&F&16~$\sim$~18\\
&~~60~$\rightarrow$~300 K: 1.0 K/minute&S&hours/cycle\\
\midrule
7th to&300~$\rightarrow$~~~~~2 K: 1.0 K/minute&S&9~$\sim$~10\\
the last&~~~~2~$\rightarrow$~300 K: 1.0 K/minute&S&hours/cycle\\
\bottomrule
\end{tabular}
\end{center}
\vspace{-2mm}
\footnotesize{
$^{*}$This recipe is based on the assumption that the PPMS is used with its protocol described in \autoref{sec:met}. Of course it is possible to find a better recipe that enables a quicker realization of the zero-resistance state within the smaller number of \emph{R--T} cycles.
\\[2pt]
$^{\dag}$Letter ``S'' and ``F'' indicates that the resistance is measured while sweeping the temperature and fixing it at each measurement point, respectively.
}
\end{table}

\section{Conclusions}
\label{sec:conc}
This study was initiated by the BCS theory, and the phonon-engineered metal sheet exhibited the resistance drop and magnetic flux expulsion at temperatures higher than 175 K. Additionally the magnetization value remained negative even at 300 K, indicating that the superconductivity survived at the temperature. Despite the used technique of `phonon' engineering, the $T_{c}$ far exceeds the BCS-McMillan prediction limit of 30's K~\cite{Mc1968}. Therefore this study can no longer rely on the BCS theory. Although engineered 2D phonons may have something to do with the activation of the 2D electron system as mentioned in the Introduction section, they cannot directly contribute to the occurrence of the high-$T_{c}$ superconductivity.

Besides the high $T_{c}$, there arose another big problem, maybe bigger than the $T_{c}$ itself. It turned out that the $T_{c}$ in a warming process was higher than that in a cooling one. In other words, the $T_{c}$ was increased with the surrounding temperature increased. This experimental fact completely disregards the BCS theory and its absolute benefit of lower temperature{\if0 for an occurrence of superconductivity\fi} that almost all superconductors are reaping.

In this study, the biggest problem was discussed on the basis of the theory of \emph{hole superconductivity}, which teaches us the absolute benefit of expansion that leads to the lowering of quantum kinetic energy. The expansion of lattice, which is likely to happen in a warming process, may account for the significant experimental fact. Of course this last remark is to be taken as indicating roughly a possible superconducting transition. But still, it is remarkable that the Hirsch's lattice expansion ultimately forms the London and Slater's \emph{giant atom} in its own framework and that the \emph{giant atom} explains the experimentally observed $H_{c}$, absurdly larger than 12 T, by its own simple \cref{eq:hc}.

Finally, the following Matthias's statement in 1973 has profound implications~\cite{Matt1973}: ``\emph{From now on, I shall look for systems that should exist, but won't -- unless one can persuade them.}'' Firstly, a high-$T_{c}$ superconductor that he was looking for does not inherently exist in this universe even at its edge, and secondly, whom he will give due credit to is the one who has actually created it, successfully persuading it to exist in this universe. A graduate student at the University of Alabama in Huntsville, Jim Ashburn, was the first to conceive of a composition that led to the discovery of YBCO~\cite{Ashburn1987}. This original formulation was simply an effort to preserve the K\textsubscript{2}NiF\textsubscript{4} structure (A\textsubscript{2}BO\textsubscript{4} formula) of La\textsubscript{1.8}Sr\textsubscript{0.2}CuO\textsubscript{4--y} by simply adjusting the relative amounts of yttrium and barium in ``Y\textsubscript{2--x}Ba\textsubscript{x}CuO\textsubscript{4--y}'' to match the average volume of the A-site ions.{\if0 Unlike attempts by other groups to make straight substitutions of lanthanum with yttrium in the La\textsubscript{1.8}Ae\textsubscript{0.2}CuO\textsubscript{4--y} (Ae: an alkali earth element) compounds,\fi} The selection of barium (over calcium and strontium) and, in turn, a substantial proportion of it did prove sufficient to produce a \emph{perovskite-like} structure, serendipitously one with remarkable properties~\cite{IEEE}.

Thus, simply doing something unique and interesting that one likes to do the best did invite serendipity~\cite{Jim2022}. As with the YBCO, the PnM was produced in that way, ignoring the dull ``science'' race towards a new world record. And thus arrived serendipity for the PnM as its ``crystal'' structure. Whether or not the study was initiated by the right theory does not matter, since ``\emph{Progress in science comes when experiments contradict theory},'' according to the Feynman's view on `science'. Although this study was initiated by the BCS theory, the initial motivation eventually turned out to be totally incompatible with the experimental results. In particular, BCS never agree with the benefit of warming to superconductivity. In accordance with another Feynman's teachings, ``\emph{It doesn't matter how beautiful your theory is, it doesn't matter how smart you are. If it doesn't agree with experiment, it's wrong},'' the BCS theory has to be shelved forever with my sincere gratitude for its help in beginning this study.
{}

It turned out that the ``crystal'' structure of the PnM was a 2D square lattice networked by oxygen (see inset of \cref{fig3}b), which is the bare essential of the high-$T_{c}$ copper--oxygen system, whereas the composite metal is not Cu but Nb and, moreover, is not in an atomic scale but in an ``island'' scale with the approximate diameter of 15 $\si{\micro m}$. According to the theory of \emph{hole superconductivity}~\cite{Hirsch1989}, a moment on the cation is not needed. That is, it does not matter whether Cu or Nb. Close approach between anions (O$^{2-}$) is important.

\begin{figure}[h!]
\centering
\includegraphics[width=0.65\linewidth]{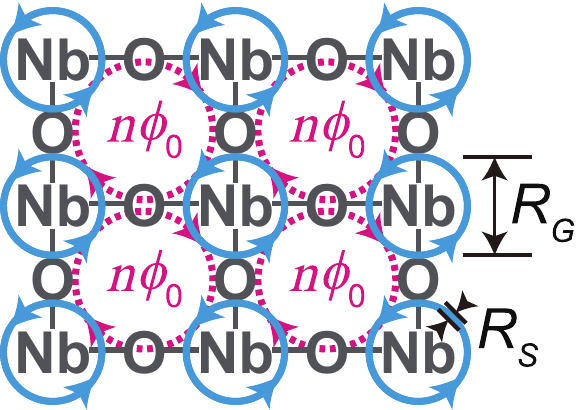}
\caption{Schematic configuration of superconducting `Nb giant atoms' with $R_{G}$ and $R_{S}$, networked by oxygen. Cyan arrow, a supercurrent generated at the moment when electrons go superconducting. Magenta arrow, the dual of the supercurrent. It circulates in the clockwise direction, thus producing anti-$n\phi_{0}$ ($n=$ 1, 2, ...) in its interior, i.e., the void of the PnM lattice.
\label{fig10}}
\end{figure}

By taking into account the \emph{real} appearance of the \emph{giant atom} in the PnM-Nb, which was proven by the large $H_{c}$ in its own right (see \cref{fig8}), the Nb ``island'' networked by oxygen may be able to be regarded as a `giant atom' with $R_{G}$ of 15 $\si{\micro m}$. From the 2D perspective, then, the superconducting coherence length $\xi$ of the PnM-Nb has to correspond to the $R_{G}$ ($\approx$ 15 $\si{\micro m}$). Additionally, the penetration depth $\lambda_{\perp}$ under a perpendicular magnetic field has to correspond to the $R_{S}$ ($\approx$ 14.5 nm). The thus relation, $\xi\gg \lambda_{\perp}$, shall indeed be consistent with the intrinsic type-I superconductivity of the PnM-Nb, namely the initial sharp resistance drop (gray curve in \cref{fig4}b). The resistive broadening observed in subsequent \emph{R--T} cycles despite the zero applied field during the measurement, on the other hand, may be accounted for by anti-$\phi_{0}$'s produced in the void each time the `Nb giant atoms' go superconducting (see \cref{fig10} and its caption). The more times the \emph{R--T} cycle is repeated, the more times the PnM-Nb undergoes the superconducting transition, therefore, the more anti-$\phi_{0}$'s the PnM lattice produces and traps in its voids. Their trapping after the multiple \emph{R--T} cycles shall change the PnM itself into the so-called type-II superconductor with the resistive transition broadened.

Besides the resistive broadening, another key feature of the resistive transition, namely the thermal hysteresis loop (blue and red curves in \cref{fig4}a for instance), can also be accounted for by the `phenomenological' Ginzburg-Landau theory. See elegant and educational ref.~\cite{HMarxiv2021Jan}. (The `dynamics' of the hysteresis loop, on the other hand, was already given in \cref{sec:expansion} in terms of the lowering of kinetic energy~\cite{HirschJAP2021,HirschBook} associated with the lattice expansion.) Thermal hysteresis is the indication of a first order phase transition, and the irreversible transition is possible if an examined specimen changes its phase during the temperature cycle. This is the case for the PnM that produces anti-$\phi_{0}$'s in its interior during the temperature cycle, changing itself from a type-I to type-II superconductor. Note in a first order phase transition only direct hysteresis is allowed. That is, the rise of resistance upon warming \emph{always} occurs at a \emph{higher} temperature than the drop of resistance upon cooling. In other words, the \emph{benefit of warming to superconductivity} that BCS never account for is warranted `phenomenologically' by the Ginzburg-Landau theory and `dynamically' by the theory of hole superconductivity. The experimental result that preserves the completeness of physics has to be taken seriously.

A proof of the anti-$\phi_{0}$'s will be shown in the forthcoming paper after the Josephson plasma frequency of the PnM lattice is estimated. Under the mixed regime of superconductivity and nonsuperconductivity, an applied current shall drive their resistive motion. Also, both the $\lambda_{\perp}$ and the extraordinarily long $\xi$, together with the critical condition that dissolves the high-$T_{c}$ superconductivity surviving at 300 K, will be shown therein.
\section{Summary}
\label{sec:summ}
To claim the occurrence of superconductivity, it is mandatory to show (i) resistance drop to zero, (ii) magnetization drop upon cooling, i.e. the Meissner effect, and (iii) the crystal structure. All the above are clearly shown in \cref{fig4}, \cref{fig5}, and \cref{fig3,fig8,fig10}, respectively. Additionally, flux trapping and the critical field are clearly confirmed at 300 K as shown in \cref{fig6} and \cref{fig7}, respectively. It is also noteworthy that all the above experimental figures are obtained directly from the experimental raw data~\cite{Zenodo2022} without any data processing such as background subtraction. Moreover, all the superconducting properties and all the superconducting transition features are fully consistent with the London-Slater-Hirsch's \emph{giant atom} and the Hirsch's theory of \emph{hole superconductivity}, respectively. Whether or not the theory is in the mainstream of science community doesn't matter. Whether or not the experimental results are correctly obtained `does' matter, whether or not the experimental results can maintain self-consistency upon the theory `does' matter.

It is interesting to see what is going to happen if Nb is substituted by other elements such as Mg, Fe, Cu. Because the used microfabrication process is not complicated, merely consisting of two deposition, one lithography and two etching, they will come soon. Also, anybody can make such PnMs since the lithography patterning file is already uploaded at Zenodo~\cite{Zenodo2022} and is freely downloadable. If they go superconducting, they will revolutionize industry and society.
\\
\\
\\
\noindent
{\small
{\bf Acknowledgments:}
The author is grateful to Masato Yamawaki for discussions, I. Shibata and S. Yeh at the Micro-Nano Open Innovation Center (MNOIC) for the HF dry etching operation, S. Suzuki and M. Akamatsu at the Nano-Processing Facility (NPF) for the XRD operation, and F. Peng and G. Fujii at the Clean Room for Analog \& Digital Superconductivity (CRAVITY) for the SEM-EDX operation.
\\
{\bf Data Availability:}
All raw data are deposited at Zenodo~\cite{Zenodo2022}, together with the GDSII file used for the sample patterning. Data are available under the terms of the Creative Commons Attribution 4.0 International license (CC BY 4.0). Addition of the author as a responsible author and/or an inventor in any publication, including electronic publications, is prohibited.
\\
{\bf Funding:}
This study was not and is not supported by any external funding. Also, this study was not and is not supported by National Institute of Advanced Industrial Science and Technology (AIST) where the author belongs, being accused of containing too much passionate motivations and conclusions to be regarded as an appropriate study for AIST. Ignoring the nonsense, the author has confidently used a part of block grant of AIST, since AIST is not the true owner of the block grant. The true owner is the citizens of Japan, to whom the author acknowledges the financial support for this study.
\\
{\bf Conflict of Interest:}
The author declares no financial/commercial conflicts of interest.
}
\urlstyle{same}


\begin{thebibliography}{99}
\bibitem{BCS1957}
  J. Bardeen, L. N. Cooper, J. R. Schrieffer, Theory of superconductivity, \href{https://doi.org/10.1103/PhysRev.108.1175}{{\it Phys. Rev.} {\bf 108} (1957) 1175--1204}.
\bibitem{Zen2014}
  N. Zen, T. A. Puurtinen, T. J. Isotalo, S. Chaudhuri, I. J. Maasilta, Engineering thermal conductance using a two-dimensional phononic crystal, \href{https://doi.org/10.1038/ncomms4435}{{\it Nat. Commun.} {\bf 5} (2014) 3435}.
\bibitem{Mal2015}
  M. Maldovan, Phonon wave interference and thermal bandgap materials, \href{https://doi.org/10.1038/nmat4308}{{\it Nat. Mater.} {\bf 14} (2015) 667--674}.
\bibitem{Soto2019}
  M. Sledzinska, B. Graczykowski, J. Maire, E. Chavez-Angel, C. M. Sotomayor-Torres, F. Alzina, 2D phononic crystals: progress and prospects in hypersound and thermal transport engineering, \href{https://doi.org/10.1002/adfm.201904434}{{\it Adv. Funct. Mater.} {\bf 30} (2019) 1904434}.
\bibitem{Nom2020}
  R. Anufriev, M. Nomura, Ray phononics: thermal guides, emitters, filters, and shields powered by ballistic phonon transport, \href{https://doi.org/10.1016/j.mtphys.2020.100272}{{\it Mater. Today Phys.} {\bf 15} (2020) 100272}.
\bibitem{JAP2021}
  T. Vasileiadis, J. Varghese, V. Babacic, J. Gomis-Bresco, D. Navarro Urrios, B. Graczykowski, Progress and perspectives on phononic crystals, \href{https://doi.org/10.1063/5.0042337}{{\it J. Appl. Phys.} {\bf 129} (2021) 160901}.
\bibitem{Pekola2021}
  J. P. Pekola, B. Karimi, {\it Colloquium}: Quantum heat transport in condensed matter systems, \href{https://doi.org/10.1103/RevModPhys.93.041001}{{\it Rev. Mod. Phys.} {\bf 93} (2021) 041001}.
\bibitem{Zen2019}
  N. Zen, Phonon-engineered Nb film as a Mott-insulating tunnel-junction network, \href{https://doi.org/10.1063/1.5126616}{{\it AIP Adv.} {\bf 9} (2019) 095023}.
\bibitem{Bed1986}
  J. G. Bednorz, K. A. M\"{u}ller, Possible high $T_{c}$ superconductivity in the Ba-La-Cu-O system, \href{https://doi.org/10.1007/BF01303701}{{\it Z. Phys. B} {\bf 64} (1986) 189--193}.
\bibitem{Ashburn1987}
  M. K. Wu, J. R. Ashburn, C. J. Torng, P. H. Hor, R. L. Meng, L. Gao, Z. J. Huang, Y. Q. Wang, C. W. Chu, Superconductivity at 93 K in a new mixed-phase Y-Ba-Cu-O compound system at ambient pressure, \href{https://doi.org/10.1103/PhysRevLett.58.908}{{\it Phys. Rev. Lett.} {\bf 58} (1987) 908--910}.
\bibitem{Muller1987}
  K. A. M\"{u}ller, M. Takashige, J. G. Bednorz, Flux trapping and superconductive glass state in La\textsubscript{2}CuO\textsubscript{4--y}:Ba, \href{https://doi.org/10.1103/PhysRevLett.58.1143}{{\it Phys. Rev. Lett.} {\bf 58} (1987) 1143--1146}.
\bibitem{Chu:BCS}
  C. W. Chu, The evolution of HTS: $T_{c}$-experiment perspectives, pp. 391--438 in: L. N. Cooper, D. Feldman (Eds.), {\it BCS: 50 Years}. World Scientific Publishing, Singapore, 2011.
\bibitem{Zenodo2022}
  N. Zen, Raw data for High temperature superconductivity arising in a metal sheet full of holes, \href{https://doi.org/10.5281/zenodo.5885550}{Zenodo, v5, doi:10.5281/zenodo.5885550} (Publication: 15 July 2022).
\bibitem{SciAdv2020}
  G. Zhang, T. Samuely, N. Iwahara, J. Ka\v{c}mar\v{c}\'{i}k, C. Wang, P. W. May, J. K. Jochum, O. Onufriienko, P. Szab\'{o}, S. Zhou, P. Samuely, V. V. Moshchalkov, L. F. Chibotaru, H.-G. Rubahn, Yu-Shiba-Rusinov bands in ferromagnetic superconducting diamond, \href{https://doi.org/10.1126/sciadv.aaz2536}{{\it Sci. Adv.} {\bf 6} (2020) eaaz2536}.
\bibitem{TinkhamBook}
  M. Tinkham, {\it Introduction to Superconductivity}, 2nd ed. Dover Publications, New York, 1996.
\bibitem{Kim1962}
  Y. B. Kim, C. F. Hempstead, A. R. Strnad, Critical persistent currents in hard superconductors, \href{https://doi.org/10.1103/PhysRevLett.9.306}{{\it Phys. Rev. Lett.} {\bf 9} (1962) 306--309}.
\bibitem{Esquin2012}
  T. Scheike, W. B\"{o}hlmann, P. Esquinazi, J. Barzola-Quiquia, A. Ballestar, A. Setzer, Can doping graphite trigger room temperature superconductivity? Evidence for granular high-temperature superconductivity in water-treated graphite powder, \href{https://doi.org/10.1002/adma.201202219}{{\it Adv. Mater.} {\bf 24} (2012) 5826--5831}.
{}
\bibitem{LondonBro1934}
  F. London, H. London, The electromagnetic equations of the supraconductor, \href{https://doi.org/10.1098/rspa.1935.0048}{{\it Proc. Roy. Soc. A} {\bf 149} (1935) 71--88}.
\bibitem{London1937}
  F. London, On the nature of the superconducting state, \href{https://doi.org/10.1103/PhysRev.51.678}{{\it Phys. Rev.} {\bf 51} (1937) 678--679}.
\bibitem{Slater1937}
  J. C. Slater, The nature of the superconducting state. II, \href{https://doi.org/10.1103/PhysRev.52.214}{{\it Phys. Rev.} {\bf 52} (1937) 214--222}.
\bibitem{LondonBro2011}
  S. Blundell, The forgotten brothers, \href{https://doi.org/10.1088/2058-7058/24/04/34}{{\it Phys. World} {\bf 24} (2011) 26--29}.
\bibitem{HirschJAP2021}
  J. E. Hirsch, Hole superconductivity xOr hot hydride superconductivity, \href{https://doi.org/10.1063/5.0071158}{{\it J. Appl. Phys.} {\bf 130} (2021) 181102}.
\bibitem{Test1974}
  L. R. Testardi, J. H. Wernick, W. A. Royer, Superconductivity with onset above 23\si{\degree}K in Nb--Ge sputtered films, \href{https://doi.org/10.1016/0038-1098(74)90002-7}{{\it Solid State Commun.} {\bf 15} (1974) 1--4}.
\bibitem{Gavaler1974}
  J. R. Gavaler, M. A. Janocko, C. K. Jones, Preparation and properties of high-$T_{c}$ Nb--Ge films, \href{https://doi.org/10.1063/1.1663717}{{\it J. Appl. Phys.} {\bf 45} (1974) 3009--3013}.
\bibitem{Matt1955}
  B. T. Matthias, Empirical relation between superconductivity and the number of valence electrons per atom, \href{https://doi.org/10.1103/PhysRev.97.74}{{\it Phys. Rev.} {\bf 97} (1955) 74--76}.
\bibitem{Matt1965}
  B. T. Matthias, T. H. Geballe, R. H. Willens, E. Corenzwit, G. W. Hull Jr., Superconductivity of Nb\textsubscript{3}Ge, \href{https://doi.org/10.1103/PhysRev.139.A1501}{{\it Phys. Rev.} {\bf 139} (1965) A1501--A1503}.
\bibitem{Matt1971}
  B. T. Matthias, The search for high-temperature superconductors, \href{https://doi.org/10.1063/1.3022880}{{\it Phys. Today} {\bf 24} (1971) 23--28}.
\bibitem{Matt1973}
  B. T. Matthias, Criteria for superconducting transition temperatures, \href{https://doi.org/10.1016/0031-8914(73)90199-7}{{\it Physica} {\bf 69} (1973) 54--56}.
\bibitem{Maple2015}
  C. W. Chu, P. C. Canfield, R. C. Dynes, Z. Fisk, B. Batlogg, G. Deutscher, T. H. Geballe, Z. X. Zhao, R. L. Greene, H. Hosono, M. B. Maple, Epilogue: Superconducting materials past, present and future, \href{https://doi.org/10.1016/j.physc.2015.03.003}{{\it Physica C} {\bf 514} (2015) 437--443}.
\bibitem{Kikoin1932}
  I. Kikoin, B. Lasarew, Hall effect and superconductivity, \href{https://doi.org/10.1038/129057b0}{{\it Nature} {\bf 129} (1932) 57--58}.
\bibitem{Born1948}
  M. Born, K. C. Cheng, Theory of superconductivity, \href{https://doi.org/10.1038/1611017a0}{{\it Nature} {\bf 161} (1948) 1017--1018}.
\bibitem{Feynman1957}
  R. P. Feynman, Superfluidity and superconductivity, \href{https://doi.org/10.1103/RevModPhys.29.205}{{\it Rev. Mod. Phys.} {\bf 29} (1957) 205--212}.
\bibitem{Chap1979}
  I. M. Chapnik, On the empirical correlation between the superconducting $T_{c}$ and the Hall coefficient, \href{https://doi.org/10.1016/0375-9601(79)90020-3}{{\it Phys. Lett. A} {\bf 72} (1979) 255--256}.
\bibitem{Hirsch1989}
  J. E. Hirsch, Hole superconductivity, \href{https://doi.org/10.1016/0375-9601(89)90370-8}{{\it Phys. Lett. A} {\bf 134} (1989) 451--455}.
\bibitem{HirschWeb}
  J. E. Hirsch, Hole Superconductivity, \url{https://jorge.physics.ucsd.edu/hole.html} (Accessed: July 2022).
\bibitem{LondonBook}
  F. London, {\it Superfluids Vol. I}. Dover Publications, New York, 1961.
\bibitem{HirschSpin2007}
  J. E. Hirsch, Spin Meissner effect in superconductors and the origin of the Meissner effect, \href{https://doi.org/10.1209/0295-5075/81/67003}{{\it Europhys. Lett.} {\bf 81} (2008) 67003}; \href{https://arxiv.org/abs/0710.0876}{arXiv:0710.0876v2} (Submission: 21 November 2007).
\bibitem{HirschBook}
  J. E. Hirsch, {\it SUPERCONDUCTIVITY BEGINS WITH H}. World Scientific Publishing, Singapore, 2020.
\bibitem{JT2022}
  A. Bussmann-Holder, H. Keller, Superconductivity and the Jahn--Teller polaron, \href{https://doi.org/10.3390/condmat7010010}{{\it Condens. Matter} {\bf 7} (2022) 10}.
{}
\bibitem{Mc1968}
  W. L. McMillan, Transition temperature of strong-coupled superconductors, \href{https://doi.org/10.1103/PhysRev.167.331}{{\it Phys. Rev.} {\bf 167} (1968) 331--344}.
\bibitem{IEEE}
  J. Ashburn, First-Hand:Discovery of Superconductivity at 93 K in YBCO: The View from Ground Zero in: {\it The Engineering and Technology History Wiki (ETHW)}, \url{https://ethw.org/First-Hand:Discovery_of_Superconductivity_at_93_K_in_YBCO:_The_View_from_Ground_Zero} (Accessed: June 2022).
\bibitem{Jim2022}
  J. Ashburn, {\it private communication}, June 2022.
{}
\bibitem{HMarxiv2021Jan}
  J. E. Hirsch, F. Marsiglio, Intrinsic hysteresis in the presumed superconducting transition of hydrides under high pressure, \href{https://arxiv.org/abs/2101.07208v2}{arXiv:2101.07208v2} (Submission: 20 January 2021).
{}
\end{thebibliography}
\end{document}